%% file: main.tex
\documentclass{article}

\input{preamble}

\title{A finite element method to compute the damping rate and frequency of oscillating fluids inside microfluidic nozzles}

\author[1]{S{\o}ren Taverniers\thanks{Corresponding author. The work reported here was performed while S. Taverniers was at Future Concepts Division, SRI International, 3333 Coyote Hill Road, Palo Alto, CA 94304, USA. \\\textit{Email address:} \texttt{sorentav@stanford.edu} (S. Taverniers).}}
\author[2]{Svyatoslav Korneev}
\author[2]{Christoforos Somarakis}
\author[2]{Morad Behandish}
\author[2]{Adrian J. Lew}

\affil[1]{Department of Energy Science \& Engineering, Stanford University, Stanford, CA 94305, USA}
\affil[2]{Future Concepts Division, SRI International, 3333 Coyote Hill Road, Palo Alto, CA 94304, USA}

\renewcommand{\shorttitle}{\textit{arXiv} Template}

\begin{document}
\maketitle

\input{abstract}

\input{motivation}

\input{problem}

\input{discretization}

\input{results}

\input{summary_outlook}

\input{appendix}

\bibliographystyle{unsrtnat}

\end{document}

%% file: preamble.tex
\pdfoutput=1
 \usepackage{arxiv}

\usepackage[utf8]{inputenc} 
\usepackage[T1]{fontenc}    

\usepackage[square,sort,comma,numbers]{natbib}
\usepackage{lineno,hyperref}
\usepackage[title]{appendix}
\usepackage{amsmath}
\usepackage{array}
\usepackage{amssymb}
\usepackage{bm}
\usepackage{algorithm}
\usepackage{algpseudocode}
\usepackage{enumerate}
\usepackage{subcaption}
\usepackage{paralist}
\usepackage{multirow}
\usepackage{color}
\usepackage{todonotes}
\usepackage{xspace}
\usepackage{doi}
\usepackage{bigdelim,booktabs,colortbl}
\usepackage{nicefrac}

\usepackage{nicematrix}
\usepackage{amsfonts}
\usepackage{url}
\usepackage{ulem}
\usepackage{authblk}

\newcommand{\fenics}{\textsf{FEniCS}\xspace}
\newcommand{\petsc}{\textsf{PETSc}\xspace}

\newcommand{\slepc}{\textsf{SLEPc}\xspace}

\newcommand{\matlab}{\textsf{MATLAB}\xspace}
\newcommand{\gmsh}{\textsf{Gmsh}\xspace}

\DeclareMathOperator*{\argmin}{arg\,min}

%% file: abstract.tex
\begin{abstract}

The computation of damping rates of an oscillating fluid with a free surface in which viscosity is small and surface tension high is numerically challenging. A typical application requiring such computation is drop-on-demand (DoD) microfluidic devices that eject liquid metal droplets, where accurate knowledge of damping rates for the least-damped oscillation modes following droplet ejection is paramount for assessing jetting stability at higher jetting frequencies. Computational fluid dynamics (CFD) simulations often struggle to accurately predict meniscus damping unless a very fine discretization is adopted, so calculations are computationally expensive. The faster alternative we adopt is to compute damping rates directly from the eigenvalues of the linearized problem. The surface tension term in Stokes or sloshing problems requires approximation of meniscus displacements, which introduces additional complexity in their numerical solution. We consider the combined effects of viscosity and surface tension, approximate the meniscus displacements, and construct a finite element method to compute the fluid’s oscillation modes. We prove that the method is free of spurious modes with zero or positive damping rates, and we implement it with Taylor-Hood elements for velocity and pressure, and with continuous piecewise quadratic elements for meniscus displacement. We verify the numerical convergence of the method by reproducing the solution to an analytical benchmark problem and two more complex examples with axisymmetric geometry. We obtain the spatial shape and temporal evolution (angular frequency and damping rate) of the set of least-damped oscillation modes in minutes, compared to days for a CFD simulation. The method’s ability to quickly generate accurate estimates of fluid oscillation damping rates makes it suitable for integration into design loops for prototyping microfluidic nozzles.

\end{abstract}


%% file: motivation.tex
\section{Motivation}
\label{sec:intro}

Droplet-based microfluidics \cite{Seemann2011} is an area of fluid dynamics with applications in droplet generation, microfabrication \cite{Seeto2022}, and manufacturing of core-shell particles \cite{Kashani2020}. In particular, microfluidic technologies are promising in additive manufacturing (AM) through drop-on-demand (DoD) 3D printing of parts, which involves pulsed generation of droplets \cite{Karampelas2017}. In DoD 3D printing, the material (e.g., a metal alloy) is first melted and then led through a nozzle to generate and  eject  droplets that are deposited onto a substrate, layer by layer, to produce the desired part. The resulting parts typically consist of hundreds of thousands or even millions of coalesced droplets. Part quality (e.g., low porosity and structural integrity) and consistency from one build to another are therefore intimately linked to  droplet jetting conditions. The more stable the generation of liquid droplets at the nozzle level is, the more consistent the deposition of these droplets on the substrate will be.

The coherency of jetting (how similar ejected droplets are) in DoD 3D printing is strongly associated with the pulsed droplet generation and ejection mechanism, as well as the dynamics of the residual material that remains in the nozzle following the ejection event. As a rule of thumb, more quiescent liquid material in the nozzle at the moment of droplet ejection leads to more coherent jetting. Therefore, the jetting coherency is strongly affected by the rate at which the kinetic energy of the liquid in the nozzle gets dissipated during the time interval between subsequent ejection events.

Recently, the significant impact of the rate of damping of an oscillating meniscus on jetting of inks and liquid metals in electrohydrodynamic \cite{Kim2018} and magnetohydrodynamic \cite{Seo2022} DoD 3D printers was highlighted. In these contributions, the role of the interplay between nozzle geometry and material properties in meniscus damping is emphasized. Specifically, \cite{Kim2018} uses a linear mass-spring-damper model developed in \cite{Stachewitz2009} to estimate the angular frequency and damping rate of small-amplitude oscillations of the ink inside the nozzle. The estimated damping rates are not compared to experimental measurements, but higher damping rates are found to be advantageous for high-frequency DoD jetting. 

\subsection{Challenges in Accurate and Efficient Computation of  Damping Rates}
Computing the damping rate using computational fluid dynamics (CFD) is  challenging in the limit of small fluid viscosity and high interfacial surface tension, as is the case for microfluidic problems \cite{Seo2022}. Detailed CFD models can have a hard time converging to a damping rate value within a reasonable tolerance, since both the meniscus dynamics and the boundary layers need to be resolved. Nicol\'{a}s \cite{Nicolas2002} and Kidambi \cite{Kidambi2009b} pointed out three sources of damping: (a) viscous dissipation in the boundary layers near the container walls and in the liquid's interior; (b) capillary damping due to meniscus effects and a moving contact line; and (c) contamination of the free surface. For a detailed review of past efforts elucidating these damping mechanisms on this problem, we refer the reader to these two papers.

CFD solvers become prohibitively expensive to use for repeated evaluation of how jetting stability changes with geometry for nozzle design optimization.

\subsection{Related Literature}
Numerous experimental, analytical and numerical studies have been conducted to analyze the dynamics of free-surface oscillations inside channels and rigid or deformable containers. Among the earliest works is the theoretical and experimental analysis by Benjamin and Scott \cite{Benjamin1970} in 1970 for an inviscid liquid's meniscus with pinned contact line in a narrow open channel, which was later extended by Graham-Eagle \cite{GrahamEagle1983} who also considered more complicated geometries. Following experimental investigations by Cocciaro et al. \cite{Cocciaro1993} into the static and dynamic properties of surface waves in cylinders making a static wall contact angle of $62^\circ$, Nicol\'{a}s \cite{Nicolas2002} computed the damping rate and frequency of the oscillation modes of capillary-gravity waves of a slightly viscous fluid in a rigid brimful cylinder with flat equilibrium meniscus. Later, he also considered a static contact angle of $62^\circ$ for both a pinned and a moving contact line \cite{Nicolas2005}. Following studies by Gavrilyuk et al. \cite{Gavrilyuk2006} and Chantasiriwan \cite{Chantasiriwan2009} of the linearized free-surface oscillations inside rigid containers using fundamental solutions, Kidambi \cite{Kidambi2009a} studied the effects of a curved equilibrium meniscus on the frequency and damping rate of free-surface oscillation modes in a rigid brimful cylinder by solving an eigenvalue problem with complex eigenvalues. 

Another line of research, driven mostly by engineering needs, involves the analysis of sloshing dynamics in partially filled containers undergoing accelerated motion. Sloshing is encountered in problems ranging from ground-supported storage tanks impacted by earthquakes to fuel tanks in aircraft or spacecraft carrying out sharp maneuvers \cite{Arafa2007}. A few authors used CFD simulations based on the volume of fluid (VoF) method to simulate liquid sloshing with and without anti-sloshing baffles, or a combination of CFD and non-CFD techniques \cite{Farhat2013}. Cruchaga et al. \cite{Cruchaga2013} performed finite element computations using a monolithic solver that included a stabilized formulation and a Lagrangian tracking technique for updating the free surface. Choudhary et al. \cite{Choudhary2021} employed analytical techniques as well as finite element and boundary element methods to study liquid oscillations in rigid circular cylindrical shells with internal flexible membranes or covered by membranes. Viola and Gallaire \cite{Viola2018} consider a theoretical approach investigating the simultaneous effects of viscous and capillary dissipation of sloshing waves in a cylindrical geometry, focusing on evaluating a contact-angle model. They  solve for an asymptotic expansion of the dynamics on a cylindrical domain through the use of Chebyshev polynomials, and  compute the wave frequency and damping rate along the way. They do account for the small changes in the domain due to the motion of the meniscus by mapping the Chebyshev polynomials from the cylinder to the deformed region.

The majority of efforts in characterizing liquid sloshing, however, were focused on linearizing the governing Navier-Stokes equations around an equilibrium meniscus and using finite elements to discretize a variational formulation of the resulting problem (\cite{Biswal2003}, \cite{Arafa2007}, \cite{Wang2006}, \cite{Miras2012}, \cite{Ohayon2015}, \cite{Ohayon2016}, and \cite{ElKamali2011}); the latter approach results in a generalized matrix eigenvalue problem that is suitable for a modal analysis yielding the eigenmodes (velocity, pressure and meniscus surface deformation profiles) and corresponding eigenvalues, yielding the modes' angular frequencies for inviscid liquids, and both their angular frequencies and damping rates for viscous liquids. These finite element-based efforts typically account for gravity, but do not simultaneously account for viscosity, surface tension, and arbitrary container geometries. They are part of a broader research thrust to deploy model order reduction techniques for performing modal analyses of complex flows, which also include proper orthogonal decomposition (POD), balanced POD, and dynamic mode decomposition (DMD), among others \cite{Taira2017}.

\subsection{Contributions}
In this work, we introduce a finite element method for computing the frequencies and damping rates of a small set of least-damped modes for fluid oscillation problems in which surface tension is important. In these problems, the oscillations of the fluid are largely driven by the potential energy cyclically stored in the surface tension of the meniscus, in contrast to typical sloshing problems, in which oscillations are mostly dominated by the gravitational potential energy. In both cases, viscous dissipation is responsible for damping, but it is the presence of the surface tension term that makes the problem numerically more complex. The finite-element method introduced here {\it simultaneously accounts for viscosity and surface tension}. While we ignore gravity and perform the linearization of the equations around a flat equilibrium meniscus to simplify the analysis, it is relatively straightforward to modify it and start from a curved meniscus instead.

On one hand, to retain a linear eigenvalue problem, it is necessary to introduce the meniscus displacement as an additional unknown, and hence to formulate a new type of mixed method in which the space of meniscus displacements needs to be appropriately selected. On the other hand, it is possible to completely eliminate the meniscus displacement from the equations, but this leads to a quadratic eigenvalue problem. In either case, a new layer  of complexity is added, one that is absent in computing modes of oscillation for the Stokes or sloshing problems. 

We formulate a general finite element method for this problem, and show that if a standard inf-sup condition for the velocity and pressure field is satisfied, and the space of meniscus displacements is included in the space of normal traces of the velocity field, then the eigenvalues of the discrete problem all have a negative real part, as those of the  exact problem do. Therefore, no modes with zero or positive damping rate arise as a result of numerical artifacts. We show that the selection of Taylor-Hood elements for the velocity and pressure fields, and continuous, piecewise quadratic elements for the meniscus displacement field, satisfies both conditions, and these are the elements we adopt in the numerical examples. While we restrict ourselves solely to the computation of the eigenmodes to analyze late-time dynamics, the fact that all eigenvalues of the discrete problem have a negative real part suggests that the spatial discretization we adopted would be stable for explicit time integration of the system as well. We leave a theoretical proof of convergence of the method for future work, but show a strong convergence of both the computed eigenvalues and corresponding eigenmodes through numerical results. Finally, the numerical tests indicate that the computation of the damping rates and oscillation frequencies from the solution of an eigensystem is orders of magnitude faster (minutes vs. $\sim$ a day) than high-fidelity CFD runs executed on comparable hardware. This reduction in computation time enables integration of the method in this work in an optimization loop for nozzle design and prototyping. CFD has been  used in the past to identify physics-based design rules \cite{Seo2022}, but its use is impractical to account for the damping rate in a computational design optimization loop; it at best allows for the exploration of a much narrower design space. 

\subsection{Paper Structure}
In Section \ref{sec:gov_eqns}, we detail the problem setup and governing set of partial differential equations, as well as the ansatz we make to obtain solutions that are oscillation modes. We demonstrate how the resulting weak form of the governing equations yields a continuous generalized eigenvalue problem (GEP). In Section \ref{sec:discretization}, we show how a finite element discretization of the domain of interest yields a matrix GEP with complex eigenvectors, and present a procedure for numerically solving this GEP using the Krylov-Schur algorithm. In Section \ref{sec:results}, we first verify our method and implementation on a planar viscous capillary wave, and then use it to predict the meniscus dynamics for a brimful cylinder and a liquid-filled container with a generic axisymmetric shape. We offer concluding remarks and a future outlook in Section \ref{sec:conclusions}.

%% file: problem.tex
\section{Problem Setup and Governing Equations}
\label{sec:gov_eqns}

In the following, we state the linearized governing equations for the motion of an incompressible fluid inside a nozzle with a free surface, directly in their non-dimensional form. To this end, we select a characteristic length $R$ to non-dimensionalize every length in the problem. The length scale then defines a characteristic time $T=\sqrt{\rho R^3/\sigma}$, speed $U=R/T=\sqrt{\sigma/(\rho R)}$, and pressure or stress $P=\sigma/R$, where $\sigma$ is the surface tension, $\rho$ is the mass density, and $\mu$ is the dynamic viscosity of the fluid.  These quantities are adopted to non-dimensionalize time,  velocity, and  pressure, respectively. The Reynolds number is defined as $\textsf{Re}=\sqrt{\rho R\sigma/\mu^2}$.

\begin{figure}
    \centering
    \includegraphics[width=0.5\linewidth]{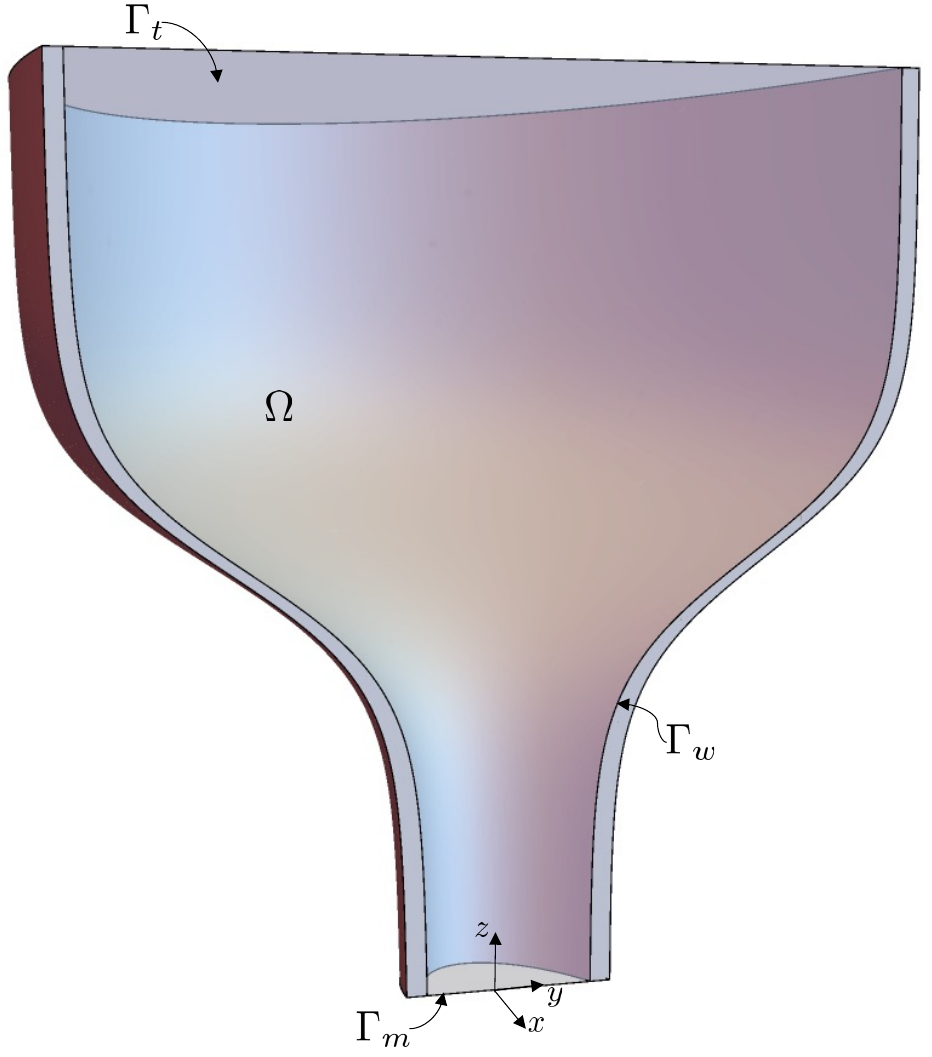}
    \caption{Sectional view of a typical nozzle design, illustrating interior and boundary domains as they are defined in \eqref{eq:linearizedequationsrealspace-nondimensional}.}
    \label{fig:domain}
\end{figure}

Consider a nozzle full of an incompressible fluid occupying an open, bounded, regular, and connected domain $\Omega\subset\mathbb R^3$ with a piece-wise smooth $2-$manifold boundary $\partial \Omega$. We partition the boundary into three disjoint surface patches; namely, the top surface $\Gamma_t$, the wall surface $\Gamma_w$, and the equilibrium meniscus surface $\Gamma_m$, each of non-zero area, such that $\partial \Omega=\overline{\Gamma_m}\cup\overline{\Gamma_w}\cup\overline{\Gamma_t}$, where $\overline{\Gamma}$ stands for the closure of $\Gamma$. Additionally, we assume that $\overline{\Gamma_m}\cap \overline{\Gamma_t}=\emptyset$, and that $\Gamma_m$'s manifold boundary is $\overline{\Gamma_m}\cap \overline{\Gamma_w}$. We consider $\Gamma_m$ to be planar and, without loss of generality, adopt a Cartesian coordinate system $O_{xyz}$ in which $\Gamma_m$ lies  in the $z=0$  plane with the unit outward normal to $\Omega$ on $\Gamma_m$ pointing towards the negative $z-$axis (see Fig. \ref{fig:domain}). The fluid inside the nozzle is assumed to be in contact with the (rigid) walls of the nozzle on $\Gamma_w$ and in contact with the atmosphere on $\Gamma_m$. We assume that $\overline{\Gamma_m} \cap \overline{\Gamma_w}$ forms a sharp edge, so that the resulting ``meniscus'' is pinned, i.e., the solid-liquid-gas contact curve remains fixed along this sharp edge. Finally, the fluid in the nozzle is in contact with more of the same fluid over the top surface $\Gamma_t$. For simplicity, we impose a stress-free condition on $\Gamma_t$, consistent with neglecting the effect of gravity in the fluid. 

We are interested in finding the late-time oscillation modes of the fluid in the nozzle, once the nonlinear convective effects become negligible and the amplitude of the meniscus oscillations is small. Under these conditions, we parameterize the surface displacement relative to $\Gamma_m$ along the $z-$axis by the function $\xi : \Gamma_m\times [0, \infty)\to \mathbb R$, i.e., $\xi(x, y, t) \in \mathbb R$ is the displacement of the meniscus at a fixed $(x,y)\in \Gamma_m$ at any time $t\geq 0$.

Let us denote the velocity field by ${\bf u}\colon \Omega\times [0, \infty)\to \mathbb R^3$, the pressure field by $p\colon \Omega\times [0, \infty)\to\mathbb R$, and the unit vector along the positive $z-$direction by ${\bf e}_z$. The linearized governing equations of fluid motion and boundary conditions are
\begin{subequations}
\label{eq:linearizedequationsrealspace-nondimensional}
    \begin{align}
        \frac{\partial {\bf u}}{\partial t} &= -\nabla p +\frac{1}{\textsf{Re}} \nabla \cdot \left( \nabla  {\bf u}+(\nabla  {\bf u})^{\sf T}\right)&& \text{in } \Omega, t > 0, \\
        \nabla \cdot {\bf u} &= 0 && \text{in } \Omega, t > 0, \\
        {\bf u} &= 0 && \text{on }\Gamma_w, t > 0, \\
        \frac{\partial \xi}{\partial t} &= {\bf u}\cdot {\bf e}_z && \text{on }\Gamma_m, t > 0, \\
        -p {\bf e}_z &+ \frac{1}{\textsf{Re}} (\nabla {\bf u}+(\nabla {\bf u})^{\sf T})\cdot{\bf e}_z = - \Delta_S \xi {\bf e}_z && \text{on }  \Gamma_m, t > 0, \label{eq:tractionBC} \\
        -p {\bf e}_z &+ \frac{1}{\textsf{Re}} (\nabla {\bf u}+(\nabla {\bf u})^{\sf T})\cdot{\bf e}_z = {\bf 0} && \text{on } \Gamma_t, t > 0, \\
        \xi &= 0 && \text{on } \overline{\Gamma_m} \cap \overline{\Gamma_w}, t > 0.
    \end{align}
\end{subequations}
where $\Delta_S$ is the surface Laplacian over $\Gamma_m$ defined by
\begin{equation}
    \Delta_S \xi \triangleq \frac{\partial^2 \xi}{\partial x^2}+\frac{\partial^2 \xi}{\partial y^2}.
\end{equation}
The external pressures on $\Gamma_t$ and $\Gamma_m$ are assumed to be the same, and set to zero here. Hence, these equations represent a linearization of the problem around a hydrostatic equilibrium at zero pressure with a flat meniscus.  Note that we do not specify any initial conditions because we are interested in the late-time dynamics of the nozzle.

To find the oscillation modes, we seek solutions to \eqref{eq:linearizedequationsrealspace-nondimensional} according to the following ansatz:%
\footnote{Or, equivalently, by applying a Laplace transform in time.}  
\begin{align}
    {\bf u}(x,y,z,t) &:= e^{\lambda t} \hat {\bf u}(x,y,z), \\
    p(x,y,z,t) &:= e^{\lambda t} \hat p(x,y,z), \\
    \xi(x,y,t) &:= e^{\lambda t} \hat \xi(x,y),
\end{align}
where $\lambda\in \mathbb C$ is a complex frequency and $\hat{\bf u}\colon\Omega\to\mathbb C^3$, $\hat{\bf p}\colon \Omega\to \mathbb C$, and $\hat \xi\colon \Gamma_m\to \mathbb C$ are unknown complex fields  for which \eqref{eq:linearizedequationsrealspace-nondimensional} becomes:
\begin{subequations}
\label{eq:linearizedequationsfourier}
    \begin{align}
    \label{eq:2}
        \lambda \hat{\bf u} &= -\nabla \hat p +\frac{1}{\textsf{Re}} \nabla \cdot \left( \nabla \hat {\bf u}+(\nabla \hat {\bf u})^{\sf T}\right) && \text{ in } \Omega, \\
        \nabla \cdot \hat {\bf u} &= 0 && \text{ in } \Omega, \label{eq:2b} \\
        \hat {\bf u} &= 0 && \text{ on }\Gamma_w, \label{eq:7} \\
        \lambda\hat \xi &= \hat {\bf u}\cdot {\bf e}_z && \text{ on }\Gamma_m,  \label{eq:2c} \\
        -\hat p {\bf e}_z &+ \frac{1}{\textsf{Re}} (\nabla \hat {\bf u}+(\nabla \hat {\bf u})^{\sf T})\cdot{\bf e}_z = -\Delta_S \hat \xi {\bf e}_z, && \text{ on }\Gamma_m \label{eq:3} \\
        -\hat p {\bf e}_z &+ \frac{1}{\textsf{Re}} (\nabla \hat {\bf u}+(\nabla \hat {\bf u})^{\sf T})\cdot{\bf e}_z = {\bf 0} &&\text{ on } \Gamma_t, \label{eq:5} \\
        \hat \xi &= 0 && \text{ on } \overline{\Gamma_m} \cap \overline{\Gamma_w}.\label{eq:4}
    \end{align}
\end{subequations}
The complex frequency $\lambda$  defines the {\it damping rate} $\eta=-\mathfrak{Re}(\lambda)$, and the {\it angular frequency} $\omega=\mathfrak{Im}(\lambda)$. 

We show in \ref{app:eigenvaluesign} that the only solution of \eqref{eq:linearizedequationsfourier} when $\lambda=0$ is the trivial one, so we henceforth only consider the case $\lambda\neq 0$. After obtaining the weak form, we show that \eqref{eq:linearizedequationsfourier} defines a generalized eigenvalue problem (GEP) with complex eigenvalues.

\subsection{Weak Form of the Problem}

The solution of problem \eqref{eq:linearizedequationsfourier} satisfies the following weak form, which is derived in \ref{app:weakform}. Let
\begin{subequations}
    \begin{align}
        \Xi & = \big\{\zeta\in H^1(\Gamma_m,\mathbb C) ~\big|~ \zeta=0 \text{ on } \overline{\Gamma_m} \cap \overline{\Gamma_w} \big\},\\
        \mathcal V & = \big\{ {\bf v}\in H^1(\Omega,\mathbb C) \times H^1(\Omega,\mathbb C) ~\big|~ {\bf v}\cdot {\bf e}_z|_{\Gamma_m} \in \Xi, {\bf v}={\bf 0} \text{ on }{\Gamma_w} \big\},\label{eq:defnu}\\
        \mathcal P & = L^2(\Omega,\mathbb C),
    \end{align}
\end{subequations}
and $v_z={\bf v}\cdot {\bf e}_z$ for ${\bf v}\in {\cal V}$. Find $\{\hat {\bf u},\hat p, \hat \xi\}\in \mathcal V\times \mathcal P\times \Xi$, $\hat {\bf u}\neq{\bf 0}$, and $\lambda\in \mathbb C$, $\lambda\neq 0$, such that for all ${\bf v}\in {\cal V}$, all $q\in {\cal P}$, and all $\zeta\in \Xi$,
\begin{subequations}
\label{eq:weaks}
    \begin{align}
        0 & = \int_{\Gamma_m}  \nabla_S\hat \xi \cdot\nabla_S v_z\; dS, \label{eq:weak1} \\
        &\qquad \qquad + \int_{\Omega}  \lambda \hat{\bf u} \cdot {\bf v}- \hat p \nabla\cdot {\bf v} + \frac{1}{\textsf{Re}} (\nabla \hat{\bf u}+(\nabla \hat{\bf u})^{\sf T})\colon \nabla{\bf v}\;dV, \nonumber \\
        0 & = \int_{\Omega} q \nabla \cdot \hat{\bf u} \; dV. \label{eq:weak2} \\
        0 & = \int_{\Gamma_m} (\nabla_S\hat u_z-\lambda \nabla_S\hat \xi)\cdot\nabla_S\zeta \; dS \label{eq:weakcompatibility}
    \end{align}
\end{subequations}
We note the weak compatibility condition between $\hat \xi$ and $\hat u_z$ expressed in terms of the equality of the gradients, instead of equality of the functions themselves. As we will see below, this is necessary to obtain a symmetric GEP. 

Notice also that we required that the space  $\mathcal V$ to contain velocity fields whose  traces on $\Gamma_m$  are in $[H^1(\Gamma_m,\mathbb C)]^2$. Some existence results require higher-regularity of the surface displacements instead (e.g., \cite{Schweizer1997,Beale1984}).  Our only consideration was  to guarantee that \eqref{eq:weakcompatibility} is well defined, and that the finite element spaces we adopt are included in these sets. This requirement removes from $\mathcal V$ some of the velocity fields in $[H^1(\Omega,\mathbb C)]^2$, whose traces on $\Gamma_m$ can be any function in  $[H^{1/2}(\Gamma_m,\mathbb C)]^2$.

The weak form can be more succinctly expressed through the following bilinear forms defined for any ${\bf u},{\bf v}\in {\cal V}$, $p,q\in{\cal P}$ and $\xi,\zeta\in \Xi$:
\begin{subequations}
\label{eq:bilinearforms}
    \begin{align}
        a({\bf {u}},{\bf v}) &= \int_{\Omega}  {\bf u} \cdot {\bf v}\;dV, \\
        b({\bf {u}},q) & = \int_{\Omega}-q\nabla \cdot {\bf {u}}\; dV, \\
        c({\bf {u}},{\bf v}) & = \int_{\Omega} \frac{1}{\textsf{Re}} (\nabla {\bf u}+(\nabla {\bf u})^{\sf T})\colon \nabla{\bf v}\; dV, \\
        s(\xi,\zeta) & = \int_{\Gamma_m}  \nabla_S \xi \cdot\nabla_S \zeta\; dS,
    \end{align}
and
    \begin{align}
        g\big(({\bf u},p,  \xi), ({\bf v},q,\zeta) \big)& =  c( {\bf u}, {\bf v})+b({\bf v}, p)+s( \xi, v_z) + b( {\bf u},q) +s( u_z, \zeta), \label{eq:defofg}\\
        h\big(({\bf u},p,  \xi), ({\bf v},q,\zeta) \big) & = -a( {\bf u}, {\bf v}) + s( \xi, \zeta).
    \end{align}
\end{subequations}
With these, equations \eqref{eq:weaks} can be restated as
\begin{subequations}
\label{eq:weakupabstract}
    \begin{align}
        0 & = \lambda a(\hat {\bf u}, {\bf v})+ c(\hat {\bf u}, {\bf v})+b({\bf v},\hat p)+s(\hat \xi, v_z), \label{eq:weaku} \\
        0 & = b(\hat {\bf u},q),\label{eq:weakp} \\
        0 & =   s(\hat u_z, \zeta)-\lambda s(\hat \xi, \zeta),
    \end{align}
or, more succinctly, as
    \begin{align}
        0 = g \big((\hat{\bf u},\hat p, \hat \xi), ({\bf v},q,\zeta) \big)-\lambda\; h \big((\hat{\bf u},\hat p,  \hat\xi), ({\bf v},q,\zeta) \big). \label{eq:symmetriceigenvaluepb}
    \end{align}
\end{subequations}
When written in this way, it is clear that the weak form defines a GEP in which $\lambda$ is the eigenvalue and $(\hat{\bf u},{\hat p},\hat \xi)$ is the eigenvector.
Setting $\hat{\bf u}={\bf 0}$, $\hat p=0$, and $\hat \xi=0$ satisfies these equations for any $\lambda\in \mathbb C$. However, we are looking for nontrivial solutions, and those are obtained for selected values of $\lambda$. Notice that both bilinear forms $g$ and $h$ are symmetric, so \eqref{eq:symmetriceigenvaluepb} defines a symmetric GEP. 

We show in \ref{app:eigenvaluesign} that all eigenvalues need to have a negative real part, i.e., $\mathfrak{Re}(\lambda)<0$, so the amplitude of each mode decays in time. Additionally, it is trivial to see that if $\lambda$ is an eigenvalue and $(\hat{\bf u},\hat p, \hat \xi)$ is its eigenvector, then its complex conjugate $\overline \lambda$ is also an eigenvalue with $(\overline{\hat{\bf u}},\overline{\hat p},\overline{\hat \xi})$ as its eigenvector. Therefore, oscillatory modes of the fluid come in pairs with frequencies of opposite signs (imaginary parts of $\lambda$ and $\overline \lambda$) and the same relaxation time (real parts of $\lambda$ and $\overline \lambda$).\footnote{Here $\overline \zeta$ indicates the complex conjugate of a complex number $\zeta \in \mathbb C$.}

In general, we are interested in finding all modes whose eigenvalue $\lambda$ satisfy $\eta \tau=-\mathfrak{Re}(\lambda)\tau<1$, for some characteristic time $\tau>0$ we choose. This corresponds to seeking all modes for which  $|\mathfrak{Re}(\lambda)|$ is smaller than $1/\tau$, i.e., which will take longer to decay than the chosen relaxation time $\tau$.

\subsection{Restriction to 2D}
\label{sec:two-dimensional}

In the following, we consider domains $\Omega$ and flows that can be represented in two dimensions, namely, either planar or axisymmetric flows. 

For planar flows, we assume that
\begin{equation}
    \Omega := \big\{ (x,y,z)\in\mathbb R^3 ~\big|~ (x,z)\in\breve\Omega, y\in (0,L) \big\},
\end{equation}
where $\breve\Omega\subset\mathbb R^2$ and $L>0$, and that $\Gamma_i=\gamma_i\times (0,L)$ with $\gamma_i\subset\partial \breve\Omega$ and $i=b,t,$ or $m$. Additionally, we assume that none of $\hat{\bf u}$, $\hat p$, or $\hat \xi$ depend on $y$, and that $\hat{\bf u}=\hat u_x {\bf e}_x+\hat u_z{\bf e}_z$ where ${\bf e}_x$ is the unit vector along the $x$-axis. 

For axisymmetric flows, let $(r,\theta,z)$ denote the standard cylindrical coordinates. We assume that
\begin{equation}
    \Omega := \big\{(r \cos\theta, r \sin\theta,z) ~\big|~ (r,z)\in\breve\Omega, \theta\in[0,2\pi) \big\}, \label{eq_axisym}
\end{equation}
where $\breve\Omega\subset [0, \infty) \times \mathbb R$ and
\begin{equation}
    \Gamma_i:=\big\{(r\cos\theta, r\sin\theta,z)~\big|~ (r,z)\in \gamma_i, \theta\in [0,2\pi) \big\}, \label{eq_2.5D}
\end{equation}
with $\gamma_i\subset \partial \breve\Omega$ and $i=b,t,$ or $m$. Additionally, we assume that none of $\hat {\bf u}$, $\hat p$, or $\hat \xi$ depend on $\theta$, and that $\hat{\bf u}=\hat u_r{\bf e}_r+\hat u_z{\bf e}_z$, where ${\bf e}_r$ is the unit vector in the direction in which $r$ grows and $\theta$ and $z$ are constants. 

To simplify the notations hereafter, we do not make a distinction between the 2D set $\breve{\Omega} \subset \mathbb R^2$, used in \eqref{eq_axisym} or \eqref{eq_2.5D}, and its embeddings in 3D as $\Omega\cap \{(x,0,z)\in \mathbb R^3\}$ or $\Omega\cap \{(r,0,z)\in \mathbb R^3\}$, respectively. Additionally, because $\hat{\bf u}$, $\hat p$, and $\hat \xi$ change only over $\breve \Omega$, we indistinctly treat them as functions over $\Omega$ or $\breve\Omega$.

\begin{figure}
    \centering
    \includegraphics[width=0.8\linewidth]{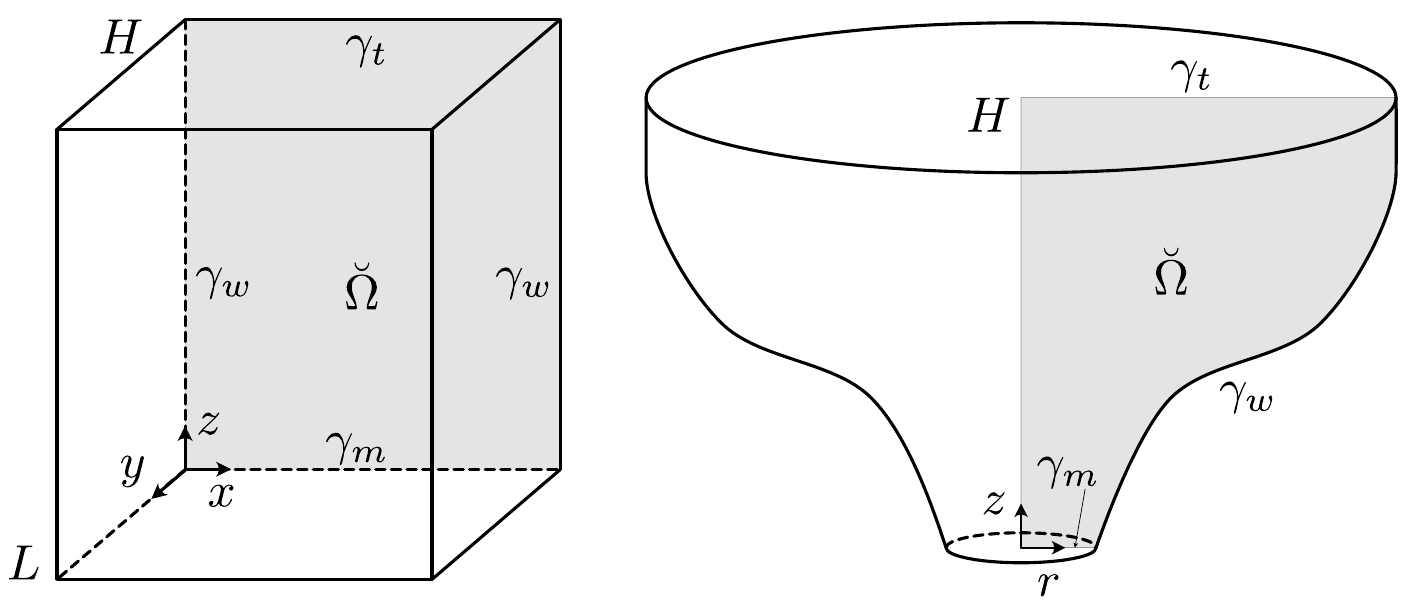}
    \caption{Domain $\breve\Omega$ for a planar (left) and an axisymmetric (right) geometry.}
    \label{fig:DomainsOfTheProblem}
\end{figure}

%% file: discretization.tex
\section{Finite Element Discretization}
\label{sec:discretization}

We discretize the 2D restriction of the problem in \S \ref{sec:two-dimensional}. We mesh $\breve\Omega$ with  triangular elements such that $\gamma_m$, $\gamma_t$, and $\gamma_w$ are each  meshed by a distinct collection of edges in the mesh, and such that every triangle has at least one node in the interior of $\breve\Omega$.%
\footnote{This is a sufficient requirement to satisfy the inf-sup condition (see \ref{app:invertibilityofG}).} On curved boundaries, a collection of edges in the mesh interpolates the boundary at the vertices, as customary. We do not implement an isoparametric construction to more accurately approximate the domain.

Over any such mesh, we construct finite element spaces $\Xi_h\subset \Xi, \mathcal V_h\subset \mathcal V$ and $\mathcal P_h\subset \mathcal P$ to approximate $\xi, {\bf v}$ and $p$, respectively. Note that by the inclusion $\Xi_h\subset \Xi$, any function $\zeta_h\in \Xi_h$ needs to satisfy $\zeta_h=0$ on $\overline{\Gamma_m}\cap \overline{\Gamma_w}$, and hence on $\overline{\gamma_m}\cap \overline{\gamma_w}$. Similarly, ${\bf v}_h\in \mathcal V_h$ needs to satisfy that ${\bf v}_h={\bf 0}$ on  $\Gamma_w$, and hence on $\gamma_w$. We will also request that
\begin{equation}
\label{eq:inclusionoftraces}
    \Xi_h\subseteq \{{\bf v}|_{\Gamma_m}\cdot {\bf e}_z\mid {\bf v}\in \mathcal V_h\}
\end{equation}
and that there exists $c>0$ such that 
\begin{equation}
    \label{eq:infsup}
    \inf_{0\not=p_h\in\operatorname{Re}(\mathcal P_{h,0})}\sup_{{\bf 0}\not={\bf v}_h\in\operatorname{Re}({\mathcal V}_{h,0})}\frac{b({\bf v}_h,p_h)}{\|{\bf v}_h\|_{1,2} \|p_h\|_{0,2}}>c>0,
\end{equation}
an inf-sup condition for the incompressibility constraint, where  $\|\cdot\|_{1,2}$ and $\|\cdot\|_{0,2}$ denote the $H^1(\Omega,\mathbb R^2)$ and $L_2(\Omega,\mathbb R)$ norms, and where ${\mathcal V}_{h,0}=\{{\bf v}\in\mathcal V_h\mid {\bf v}|_{\partial \Omega}={\bf 0}\}$ and $\mathcal P_{h,0}=\{p\in\mathcal P_h\mid \int_{\Omega} p\;dV=0\}$. 

As a concrete example, we construct the following finite element spaces for {\it planar flows}:
\begin{displaymath}
    \begin{aligned}
        \Xi_h &= \big\{\zeta\in \Xi ~\big|~ \zeta|_e\in P_2(e,\mathbb C),\text{ in all mesh edges $e$ on $\gamma_m$} \big\}, \\
        \mathcal V_h & = \big\{{\bf v}\in \mathcal V ~\big|~  (v_x,v_z)|_K\in P_2(K,\mathbb C) \times P_2(K,\mathbb C),\text{ in all mesh elements $K$ in $\breve\Omega$} \big\}, \\
        \mathcal P_h& = \big\{p\in\mathcal P ~\big|~ p|_K\in P_1(K,\mathbb C),\text{ in all mesh elements $K$ in $\breve\Omega$} \big\}.
    \end{aligned}
\end{displaymath}
Here, $P_k(D,\mathbb C)$ is the space of complex-valued polynomials of degree $k\in \mathbb N$ over the domain $D\subset \mathbb R^2$. Because of the planar flow assumption, the velocity is ${\bf v}=v_x{\bf e}_x+v_z{\bf e}_z$ in $\Omega$ for ${\bf v}\in\mathcal V_h$ and it is enough to define the fields over $\breve\Omega$, since they are extended to $\Omega$ noting that $\zeta\in \Xi_h$, $p\in \mathcal P_h$, and ${\bf v}\in \mathcal V_h$ do not depend on $y$. The construction of finite element spaces for {\it axisymmetric flows} is obtained in exactly the same way, after replacing $x$ by $r$ and $y$ by $\theta$ in the definition above. Notice that, by construction, ${\bf v}|_{\Gamma_m}\cdot{\bf e}_z\in \Xi_h$ for any ${\bf v}\in \mathcal V_h$, so \eqref{eq:inclusionoftraces}
 is satisfied.

\begin{figure}[htb]
    \centering
    \includegraphics[width=0.8\linewidth]{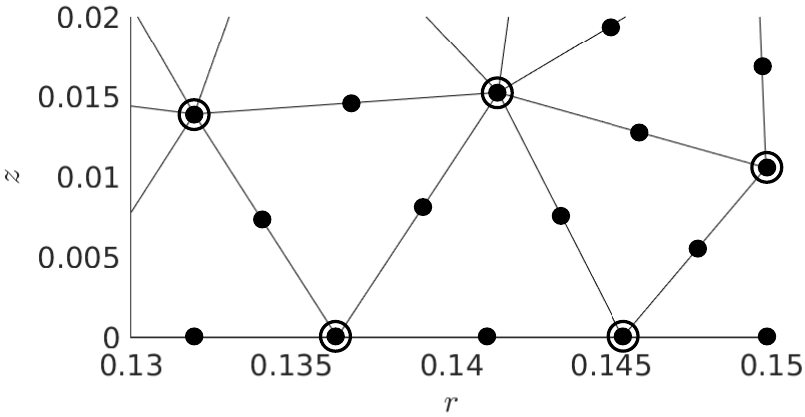}
    \caption{Section of the computational domain for the axisymmetric geometry considered in \S \ref{subsec:arbitrary_shape}. The triangles are Taylor-Hood finite elements, with filled circles indicating velocity nodes and empty circles denoting pressure nodes.}
    \label{fig:TaylorHood}
\end{figure}

The spaces $\mathcal V_h$ and $\mathcal P_h$ for the approximation of  the velocity and pressure fields, respectively, are constructed with  Taylor-Hood elements \cite{Taylor1973,Boffi1997} (Fig. \ref{fig:TaylorHood}), and they use continuous piecewise quadratic functions for each velocity component and continuous piecewise linear functions for the pressure. Taylor-Hood elements are  standard finite elements used in the simulation of incompressible fluid flow, since under mild conditions on the mesh, they satisfy the inf-sup stability condition for the incompressibility constraint \eqref{eq:infsup} in both planar \cite{LarsonBengzon2013,ern2004theory} and axisymmetric \cite{lee2011stability} flows.

The numerical approximation is obtained from the weak form \eqref{eq:weaks} by substituting the infinite-dimensional function spaces $\Xi$,  $\mathcal V$, and $\mathcal P$ by their finite-dimensional counterparts $\Xi_h$,  $\mathcal V_h$, and $\mathcal P_h$, respectively. We seek $(\hat{\bf u}_h, \hat p_h, \hat \xi_h)\in \mathcal V_h\times \mathcal P_h \times  \Xi_h$ and $\lambda\in \mathbb C$ such that 
\begin{align}
    \label{eq:symmetriceigenvaluepb_FE}
    0 =     g \big((\hat{\bf u}_h,\hat p_h, \hat \xi_h), (\hat{\bf v}_h,\hat q_h,\hat \zeta_h) \big)-\lambda\;
    h \big((\hat{\bf u}_h,\hat p_h,  \hat\xi_h), (\hat{\bf v}_h,\hat q_h,\hat\zeta_h) \big),
\end{align}
for all $(\hat{\bf v}_h, \hat q_h, \hat \zeta_h)\in \mathcal V_h\times \mathcal P_h \times  \Xi_h$. Henceforth,  we will refer to the approximate solutions computed with this finite element discretization as the finite element analysis (FEA) solutions.

We show in Appendix \ref{app:invertibilityofG} that the eigenvalues of \eqref{eq:symmetriceigenvaluepb_FE} also have a negative real part, which guarantees that the method is free of spurious modes with zero or positive damping rates. The inf-sup condition that the Taylor-Hood elements satisfy, and the choice of discrete spaces so that $\Xi_h\subseteq 
\{{\bf v}_h|_{\Gamma_m} \cdot {\bf e}_z \text{ for } {\bf v}_h\in\mathcal V_h\}$, are needed to prove this result.

\subsection{Matrix Formulation}

Problem \eqref{eq:symmetriceigenvaluepb_FE} leads to a symmetric matrix GEP. To this end, we let $\{\varphi_\alpha\}_{\alpha \in \mathsf{A}}$ denote the standard basis functions of $P_2$-elements, where index $\alpha \in \mathsf{A}$ runs over all corner and mid-nodes in the mesh except for those on $\Gamma_w$,  and $\{\psi_\beta\}_{\beta \in \mathsf{B}}$ denotes the standard basis functions for $P_1$-elements, where index $\beta \in \mathsf{B}$ runs over all corner nodes in the mesh. We can thus write $(\hat{\bf u}_h, \hat p_h, \hat \xi_h)\in \mathcal V_h\times \mathcal P_h \times  \Xi_h$ as
\begin{displaymath}
    \begin{aligned}
        \hat{\bf u}_h  = \sum_{\alpha \in \mathsf{A}} {\bf v}_\alpha \varphi_\alpha, \quad
        \hat p_h  = \sum_{\beta \in \mathsf{B}} p_\beta \psi_\beta, \quad
        \hat \xi_h  = \sum_{\alpha \in \mathsf{A}_m} \xi_\alpha \varphi_\alpha,    
    \end{aligned}
\end{displaymath}
where $\mathsf{A}_m \subset \mathsf{A}$ indicated  all node indices $\alpha \in \mathsf{A}$ that are located inside $\Gamma_m$, excluding those on the boundary $\overline{\Gamma_m} \cap \overline{\Gamma_w}$. Here ${\bf v}_\alpha\in \mathbb C^2$, $p_\beta\in \mathbb C$, and $\xi_\alpha\in\mathbb C$  for each $\alpha$ and $\beta$, and they can be grouped in column matrices $U$, $P$, and $Z$, respectively. 
The symmetric GEP can then be stated as finding $X=[U\ P\ Z]^{\sf T}$ and $\lambda\in \mathbb C$ such that
\begin{subequations}
    \begin{equation}
        \label{G_H_GEP_detailed}
        \underbrace{\begin{bmatrix}
            C & B & S \\ B^{\sf T} & 0 & 0 \\ S&0&0 
        \end{bmatrix}}_{G}
        \begin{bmatrix}
            U \\ P \\ Z 
        \end{bmatrix}
        =\lambda  
        \underbrace{ \begin{bmatrix}
        -A & 0 & 0\\ 0 & 0 & 0\\ 0 & 0 & S    
        \end{bmatrix}}_{H}
        \begin{bmatrix}
            U \\ P \\ Z 
        \end{bmatrix}    
    \end{equation}
    or
    \begin{equation}
        \label{G_H_GEP}
        G X = \lambda H X,
    \end{equation}
\end{subequations}
where the elements of the matrices $A, B, C,$ and $S$ are defined as
\begin{align*}
    A_{\alpha, \alpha'} &=a(\varphi_{\alpha}, \varphi_{\alpha'}), \quad \alpha, \alpha' \in \mathsf{A}, \qquad
    B_{\alpha, \beta} =b(\varphi_\alpha, \psi_\beta), \quad \alpha \in \mathsf{A}, \beta \in \mathsf{B}, \\
    C_{\alpha, \alpha'} &=c(\varphi_{\alpha}, \varphi_{\alpha'}), \quad \alpha, \alpha' \in \mathsf{A}, \qquad
    S_{\alpha, \alpha'} =s(\varphi_{\alpha}, \varphi_{\alpha'}), \quad \alpha, \alpha' \in \mathsf{A}_m,
\end{align*}
in which the bilinear forms $a$, $b$, $c$, and $s$ were defined in \eqref{eq:bilinearforms}.

We show in \ref{app:invertibilityofG} that $\mathfrak{Re}(\lambda)<0$,  and hence that matrix $G$ is invertible. Therefore, the method is free from spurious modes with zero or positive damping rate, which could arise as an artifact of the discretization. The size of the system in \eqref{G_H_GEP} is given by the total number of degrees of freedom for the velocity components, pressure, and meniscus surface displacement in $\breve \Omega$, hereafter denoted by $n$.

Since $G$ and $H$ are real symmetric matrices, if $\lambda$ is a complex eigenvalue with eigenvector $X$, then $\overline{\lambda}$ is also an eigenvalue with eigenvector $\overline X$. Such a symmetry also implies that if $\lambda\in \mathbb R$, then $X$ is a real eigenvector. Each pair of complex eigenvectors $X$ and $\overline{X}$ defines complex-valued evolutions in time $X\exp(\lambda t)$ and $\overline{X}\exp(\overline\lambda t)=\overline{X \exp(\lambda t)}$. A 2D subspace of real-valued velocities, pressures, and meniscus surface displacements follows as linear combinations of   $X_+(t)=\mathfrak{Re}(X\exp(\lambda t))$ and $X_-(t)=\mathfrak{Im}(X\exp(\lambda t))$. Explicitly, if $\lambda =-\eta+i\omega$, then
\begin{displaymath}
    \begin{aligned}
        X_+(t) & = e^{-\eta t}\big(\mathfrak{Re}(X) \cos(\omega t) - \mathfrak{Im}(X) \sin(\omega t)\big), \\
        X_-(t) & = e^{-\eta t}\big(\mathfrak{Re}(X) \cos(\omega t-\pi/2) - \mathfrak{Im}(X) \sin(\omega t-\pi/2)\big).
    \end{aligned}
\end{displaymath}
Both $X_+$ and $X_-$  have the same damping rate $\eta$ and angular frequency $\omega$, but are out of phase with each other.

\subsection{Implementation and Solution of the Matrix GEP}
We implemented the finite element discretization in the open-source framework \fenics \cite{Langtangen2016}, which provides matrices $G$ and $H$ of the GEP in  \eqref{G_H_GEP}. We used built-in meshing tools from \fenics (for the planar and cylindrical domains) and \gmsh \cite{Geuzaine2009} (for the generic axisymmetric domain) to generate the finite element mesh. 

Given that we are only interested in a small subset of  pairs of eigenvalues and eigenvectors (i.e., those with the smallest values of $|\mathfrak{Re}(\lambda)|$), an iterative solver that converges only a selected set of eigenpairs is most appropriate. To this end, we used the Krylov-Schur method \cite{Stewart2001,Stewart2002,Hernandez2007}, a projection-based algorithm that uses a Krylov subspace, with a random vector to initialize this subspace. Krylov-Schur is an improvement over the traditional Arnoldi and Lanczos schemes in that it incorporates an effective and robust restarting technique, generally resulting in improved convergence. Given that $H$ is singular, in order to apply Krylov-Schur we needed to perform a shift-and-invert transform (see \cite{Roman2022} for details). 

The matrix $G$ can display poor conditioning which leads to problems with the convergence of some components of the eigenvectors. To address this problem, we left-multiplied both sides of \eqref{G_H_GEP} by a diagonal preconditioner $\Lambda$ with diagonal components given by $\Lambda_{ii}=1/\max_{j=1,\ldots, n}|G_{ij}|$  for $i=1,\ldots,n$, where $n$ is the dimension of the GEP \eqref{G_H_GEP}. This operation does not change the eigenvectors or eigenvalues of \eqref{G_H_GEP}, but it does transform the matrices of the system into non-symmetric ones. It is possible to keep the symmetry by preconditioning from the right as well, but we did not do it here.

For the computations in Section \S \ref{sec:results}, the size $n$ of the GEP \eqref{G_H_GEP} can approach $\mathcal{O}(10^6)$. It can be seen from \eqref{G_H_GEP_detailed} that $G$ and $H$ are highly sparse matrices, and thus can be efficiently represented using \petsc \cite{Balay2022} to overcome memory limitations. In turn, this enabled us to deploy the \petsc-based eigensolver \slepc \cite{Roman2022}, which has the Krylov-Schur algorithm. Additionally, we  used {\sf MUMPS} \cite{Amestoy2011} for LU factorization of large sparse matrices. The combination of these techniques resulted in a highly parallelizable eigensolver for this problem.

In \slepc, we can set flags to control the number of requested eigenvalues, denoted by \texttt{nev}, and to set the convergence criteria, i.e., maximum number of iterations to perform $N_{iter,max}$ and the tolerance $\epsilon$ below which we would like the error of the computed eigenvalues to fall. We set $N_{iter,max}=5000$ and $\epsilon=10^{-16}$, which we found to be appropriate for our problems of interest. The maximum dimension of the Krylov subspace has a default value of twice \texttt{nev}, which we used in our computations. Typically, we only needed to request \slepc to converge \texttt{nev} $=70$ eigenvalues to obtain the least-damped modes of interest. Finally, we note that the use of the  governing equations in their dimensionless form in \eqref{eq:linearizedequationsrealspace-nondimensional} was critical to obtain significantly improved robustness of the eigensolver with respect to the choice of the shift value.

\paragraph{Accounting for the Dirichlet boundary conditions on $\Gamma_w$} In order to implement the boundary conditions on $\Gamma_w$ (slip or no-slip), it is sufficient to zero out the rows of $G$ and $H$ corresponding to the constrained nodal degrees of freedom on $\Gamma_w$, and set the diagonal components of $G$ in these rows to be 1. To keep the symmetry of $G$ and $H$ it would have been necessary to zero out the corresponding columns, but because we only used a left preconditioner, we did not do this and worked with the non-symmetric matrices.

%% file: results.tex
\section{Results}
\label{sec:results}

We showcase the performance of the method and compute the least-damped modes of oscillation of a fluid in a nozzle, including the accuracy and convergence rate of the FEA solutions. To this end, we perform a suite of numerical tests aimed at verifying the code and the method on an analytical benchmark problem (planar capillary wave), and analyzing the convergence as the mesh size is decreased on cylindrical and arbitrary axisymmetric domains. 

\subsection{Analytical Benchmark Problem: 2D Inviscid and Viscous Capillary Wave}
\label{subsec:analytical_benchmark}

In this example, we consider a planar flow problem that involves oscillations of a capillary wave in a 2D $xz-$domain $\breve{\Omega} = (0,1) \times (0,H)$, for some $H\gg 1$, with $\gamma_m=(0,1)\times\{0\}$, $\gamma_w= \big( \{0,1\}\times(0,H) \big) \cup \big( [0,1]\times\{H\} \big)$, and $\gamma_t=\emptyset$.  To obtain an analytical solution, we first consider the inviscid case and find the modes of oscillation. We then use these solutions to approximate the damping rate and angular frequency for each of the modes when the viscosity is small but not zero.

\subsubsection{Inviscid Case}

When there is no viscosity and the flow is assumed to be irrotational, the system of equations \eqref{eq:linearizedequationsfourier} can be restated in terms of a velocity potential $\phi\colon{\Omega}\to \mathbb R$ so that $\nabla \phi=\hat{\bf u}$, taking advantage of the property that $\phi$  can be chosen to satisfy $p+\partial \phi/\partial t=0$ in $\Omega$ at all times $t > 0$ \cite{LandauLifshitz1987}. Since the flow is planar, neither $\xi$ nor $\phi$  depend on $y$, and using the ansatz $\phi(x,z,t)=\hat\phi(x,z)\exp(\lambda t)$ we can write $\hat p+\lambda \hat \phi=0$ in lieu of \eqref{eq:2}.
In terms of $\hat \phi$ and $\hat \xi$, \eqref{eq:2b} to \eqref{eq:5} become
\begin{subequations}
\label{problem_phi_only}
    \begin{align}
        \Delta \hat \phi & = 0, && (x, z) \in \Omega, \label{eq:cap1} \\
        \frac{\partial\hat \phi}{\partial x}\Big|_{(0,z)}= \frac{\partial\hat \phi}{\partial x}\Big|_{(1,z)} &= 0, && z\in (0, H), \label{eq:cap2} \\
         \frac{\partial\hat \phi}{\partial z}\Big|_{(x,H)} & = 0, &&  x\in (0, 1), \label{eq:cap3} \\
         \lambda \hat \xi - \frac{\partial\hat \phi}{\partial z}\Big|_{(x,0)} &=0, && x\in (0, 1), \label{eq:cap5} \\
        \lambda \hat \phi\Big|_{(x,0)} + \frac{\partial^2 \hat \xi}{\partial x^2} &= 0, && x\in (0, 1). \label{eq:cap6}
    \end{align}
\end{subequations}
Because it is inviscid flow, the no-slip boundary condition \eqref{eq:7} was replaced by the no-penetration conditions \eqref{eq:cap2} and \eqref{eq:cap3}. Also, since it is potential flow, we cannot impose \eqref{eq:4}. 

We can seek exact solutions of the form 
\begin{equation}
\label{eq:ansatzexactsolution}
    \hat \phi_n(x,z) = e^{-n \pi z}\cos(n \pi x), \qquad \hat \xi_n(x) = -\frac{n\pi}{\lambda_n}\hat\phi_n(x,0),    
\end{equation}
for $n\in \mathbb N$, since in this way \eqref{eq:cap1}, \eqref{eq:cap2}, \eqref{eq:cap5} and \eqref{eq:cap6} are satisfied. From  \eqref{eq:cap6}, it follows that
\begin{displaymath}
    \lambda_n =\pm i (n \pi)^{3/2},
\end{displaymath}
and from $\hat p+\lambda \hat \phi=0$, we obtain that
\begin{displaymath}
    \hat p(x,z) = \mp i (n\pi)^{3/2} \hat \phi_n(x,z).
\end{displaymath}
Clearly, \eqref{eq:ansatzexactsolution} cannot satisfy \eqref{eq:cap3}. However, if  $H\gg 1$, then $\frac{\partial\hat\phi_n}{\partial z}(x,H)\approx 0$ and \eqref{eq:cap3} is approximately satisfied. We use $H=4$ in our numerical experiments.

The time evolution of these capillary waves is then given by
\begin{equation}
    \begin{aligned}
        \phi_n^\pm(x,z,t) & = \cos(k_n x) \exp(-k_n z\pm i \omega_n t),\\ \xi^\pm(x,t)& =\pm i \omega_n^{-1/3} \phi_n^\pm(x,0,t),\\
        p^\pm(x,z,t) & = \mp i \omega_n \phi_n^\pm(x,0,t),   
    \end{aligned}
\end{equation}
where 
\begin{equation}
\label{eq:analyticalfrequency}
    k_n = n\pi, \qquad \omega_n=k_n^{3/2}=(n \pi)^{3/2}.
\end{equation}
The velocity field $\bf u_n^{\pm}$ follows as the spatial gradient of $\phi_n^{\pm}$.

\textcolor{cyan}{It follows from \eqref{eq:analyticalfrequency} that the expected dimensionless angular frequencies for the three lowest angular-frequency modes (i.e, $n=1, 2, 3$) are $\omega_1=5.5683$, $\omega_2=15.7496$ and $\omega_3=28.9339$. To verify the accuracy of the FEA solutions in the inviscid limit, we compute FEA solutions with progressively smaller values of dynamic viscosity, or progressively larger values of the Reynolds number \textsf{Re}.\footnote{\textcolor{cyan}{The FEA formulation needs to be modified for a zero viscosity, since there is a slip boundary condition on the boundary in this case.}} The expectation is that the angular frequencies computed with FEA will converge to their analytical counterparts \eqref{eq:analyticalfrequency} in the limit as \textsf{Re} $\to \infty$.}

\begin{figure}
\includegraphics[width=0.8\linewidth]{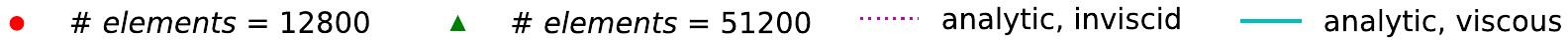}
     \centering
     \begin{subfigure}[b]{0.32\textwidth}
         \centering
         \includegraphics[width=\textwidth]{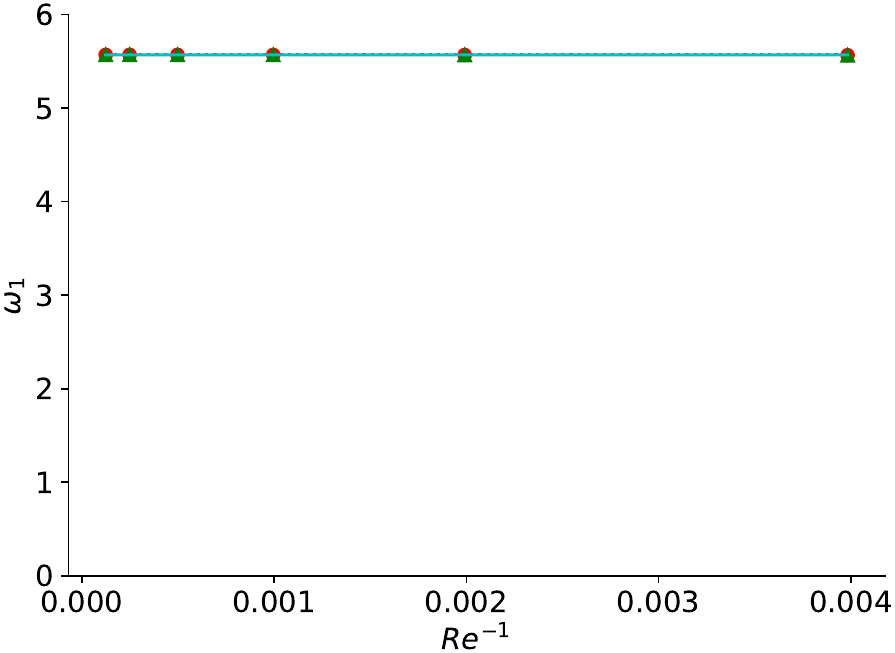}
         \caption{Mode $n=1$ (scale)}
     \end{subfigure}
     \hfill
     \begin{subfigure}[b]{0.32\textwidth}
         \centering
         \includegraphics[width=\textwidth]{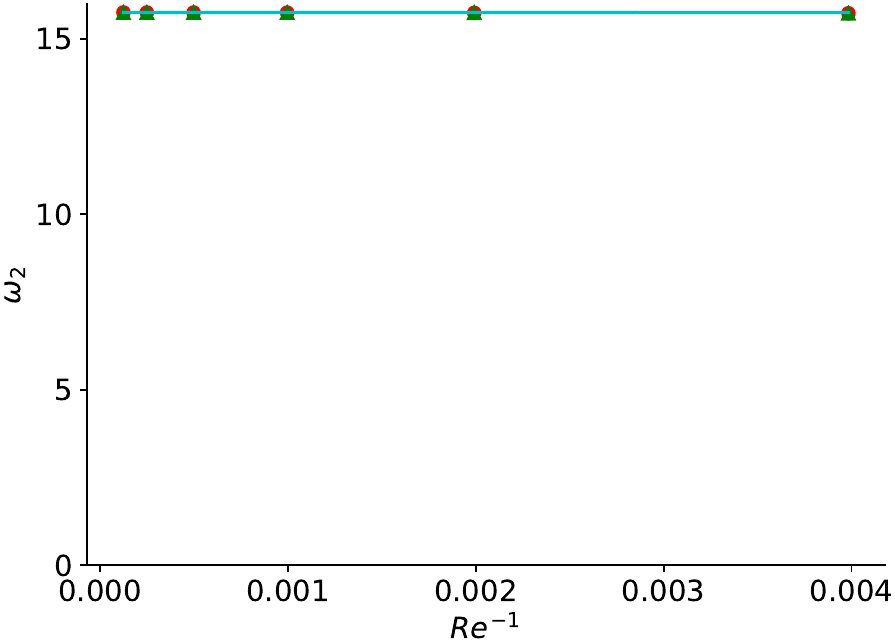}
         \caption{Mode $n=2$ (scale)}
     \end{subfigure}
     \hfill
     \begin{subfigure}[b]{0.32\textwidth}
         \centering
         \includegraphics[width=\textwidth]{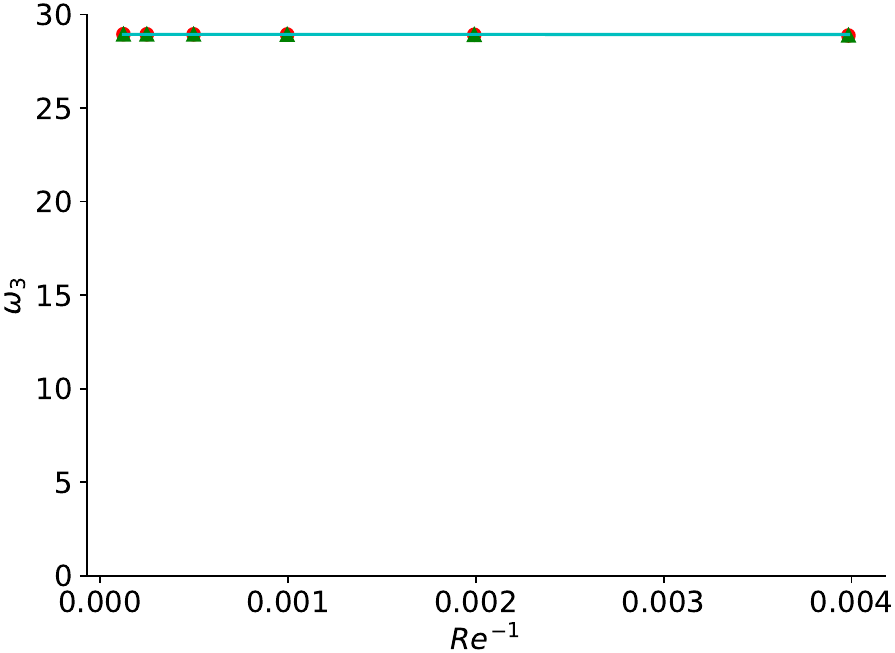}
         \caption{Mode $n=3$ (scale)}
     \end{subfigure}
     \begin{subfigure}[b]{0.32\textwidth}
         \centering
         \includegraphics[width=\textwidth]{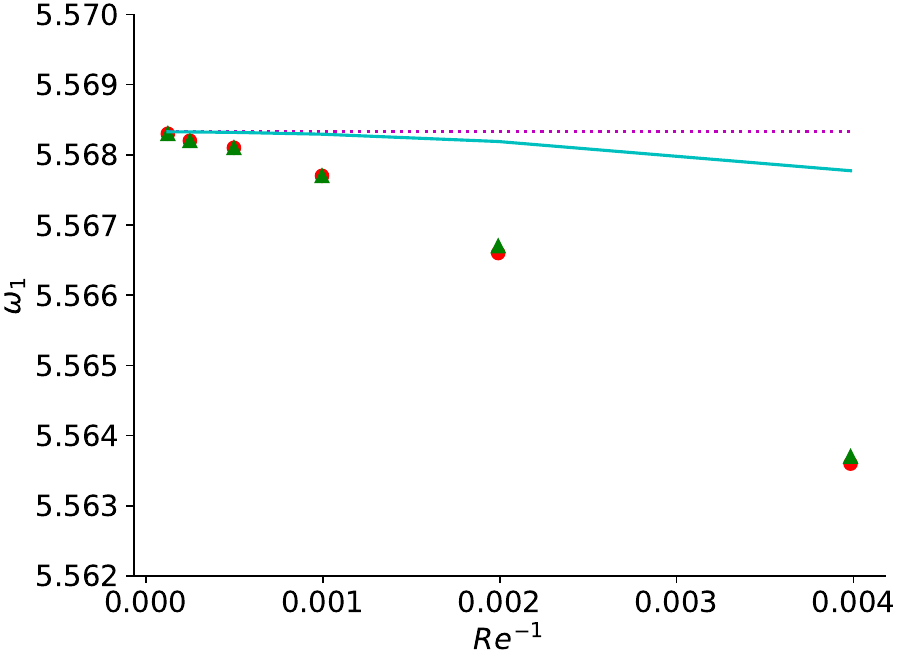}
         \caption{Mode $n=1$ (detail)}
     \end{subfigure}
     \hfill
     \begin{subfigure}[b]{0.32\textwidth}
         \centering
         \includegraphics[width=\textwidth]{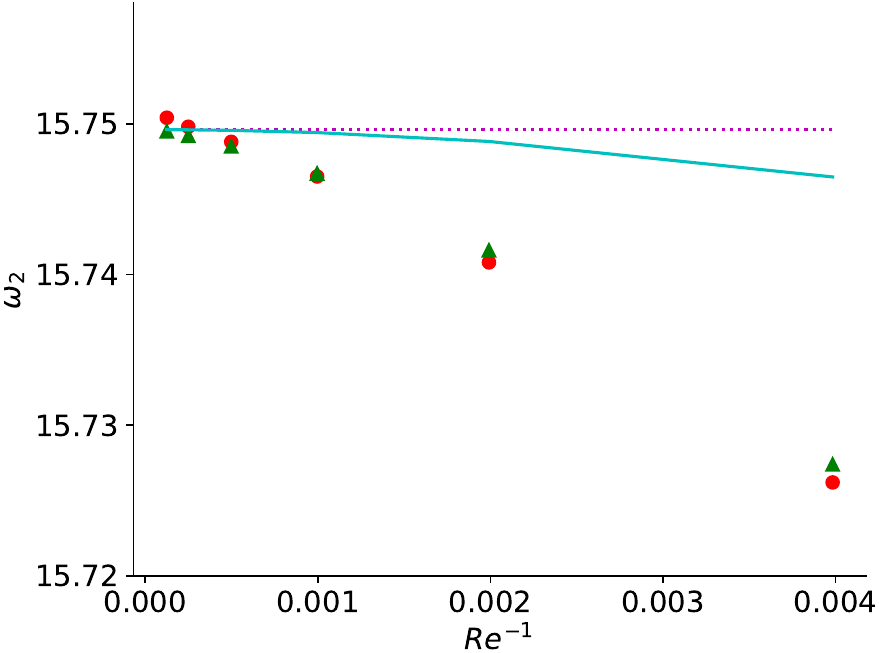}
         \caption{Mode $n=2$ (detail)}
     \end{subfigure}
     \hfill
     \begin{subfigure}[b]{0.32\textwidth}
         \centering
         \includegraphics[width=\textwidth]{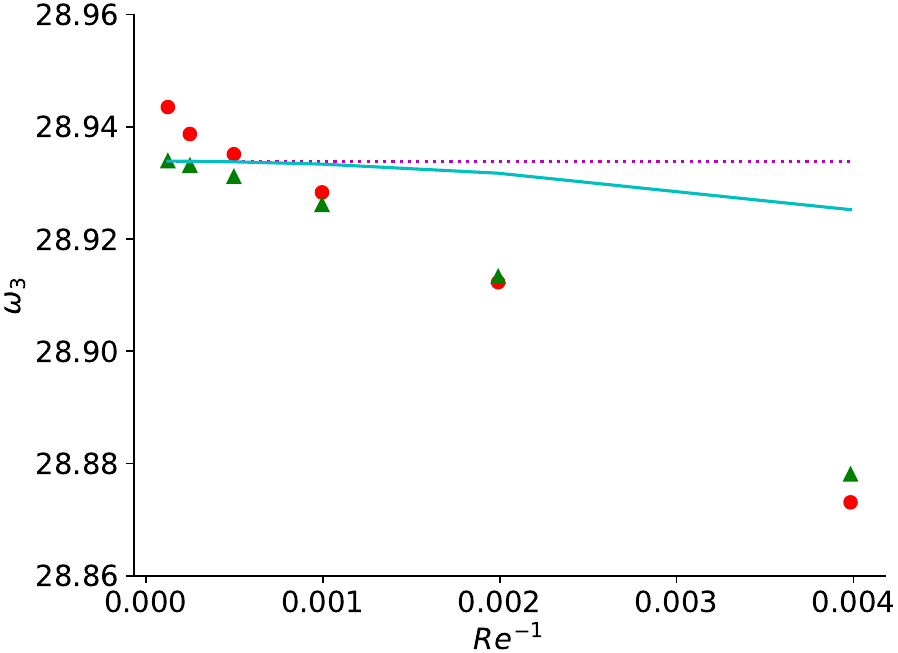}
         \caption{Mode $n=3$ (detail)}
     \end{subfigure}
     \caption{\textcolor{cyan}{Comparison of the angular frequencies of a viscous capillary wave solution obtained with both  FEA and  analytically, i.e., \eqref{eq:lambda_viscous}. Also included are the values for an inviscid capillary wave, given by \eqref{eq:analyticalfrequency}. The comparison is shown for the three lowest-angular-frequency modes:  (a) mode $n=1$, (b) mode $n=2$, and (c) mode $n=3$. The top row provides a sense of scale, highlighting the fact that the discrepancy between the FEA computed and analytical viscous values remains small (below 0.1\%) even at low Reynolds numbers. The bottom row provides a more detailed view, highlighting both the convergence of the FEA computed values to the analytical values as the Reynolds number grows, and the fact that the FEA computed values are largely mesh-converged, as a comparison between the two levels of refinement (i.e., number of finite elements) suggests. The inviscid values in red are of course constant with Reynolds number, and confirm that both viscous solutions approach them as the Reynolds number grows.}}
     \label{fig:modes_frequencies}
\end{figure}

\textcolor{cyan}{Figure \ref{fig:modes_frequencies} demonstrates that the angular frequencies for modes $n=1, 2, 3$ obtained from FEA solutions for Reynolds numbers ranging from 251 to 8034 converge to their counterparts computed with \eqref{eq:analyticalfrequency} in the limit of $\textsf{Re}^{-1}$ approaching zero.}

\begin{figure}
     \centering
     \begin{subfigure}[b]{0.45\textwidth}
         \centering
         \includegraphics[width=\textwidth]{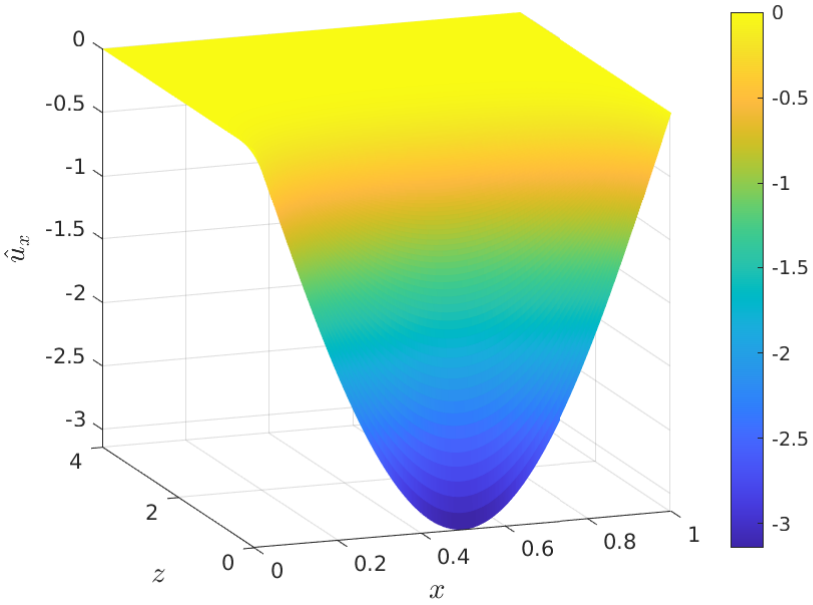}
         \caption{Analytical values of $\hat u_x$.}
     \end{subfigure}
     \begin{subfigure}[b]{0.45\textwidth}
         \centering
         \includegraphics[width=\textwidth]{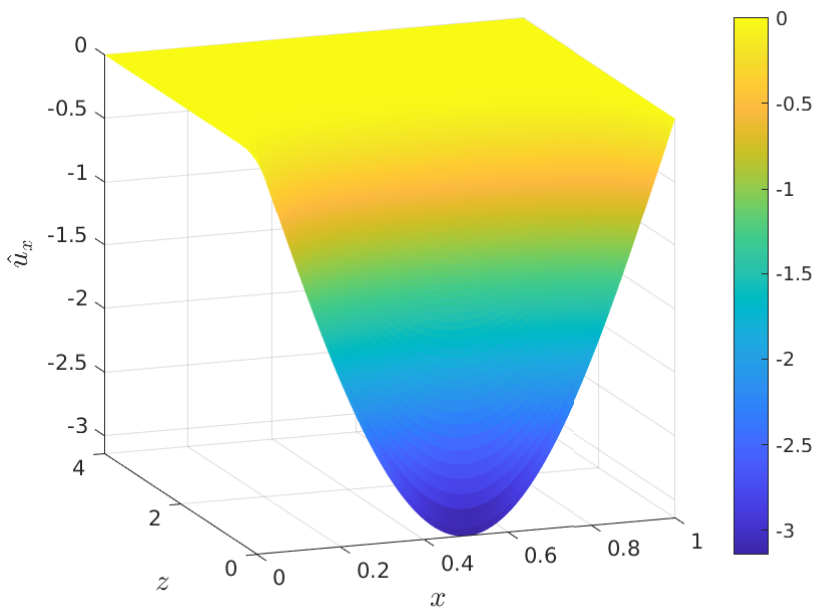}
         \caption{Computed values of $\hat u_x$ from the FEA solution.}
     \end{subfigure}
     \begin{subfigure}[b]{0.45\textwidth}
         \centering
         \includegraphics[width=\textwidth]{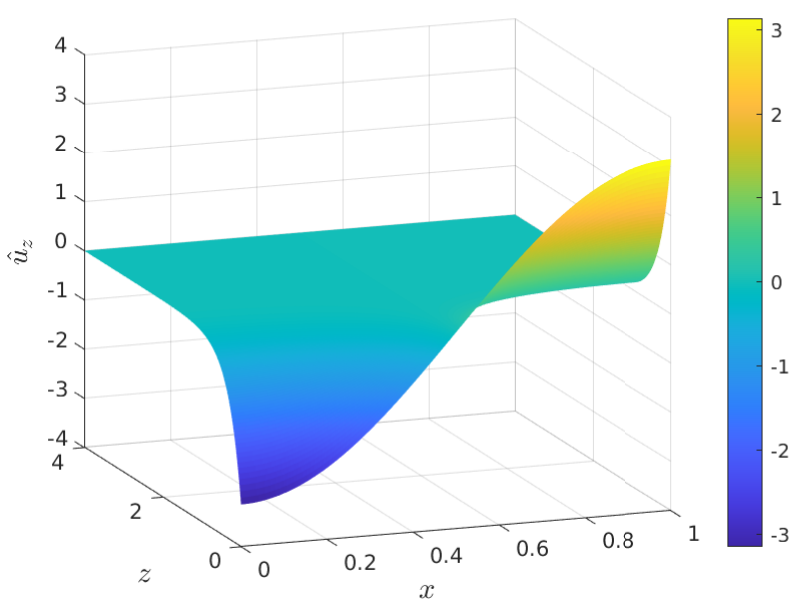}
         \caption{Analytical values of $\hat u_z$.}
     \end{subfigure}
     \begin{subfigure}[b]{0.45\textwidth}
         \centering
         \includegraphics[width=\textwidth]{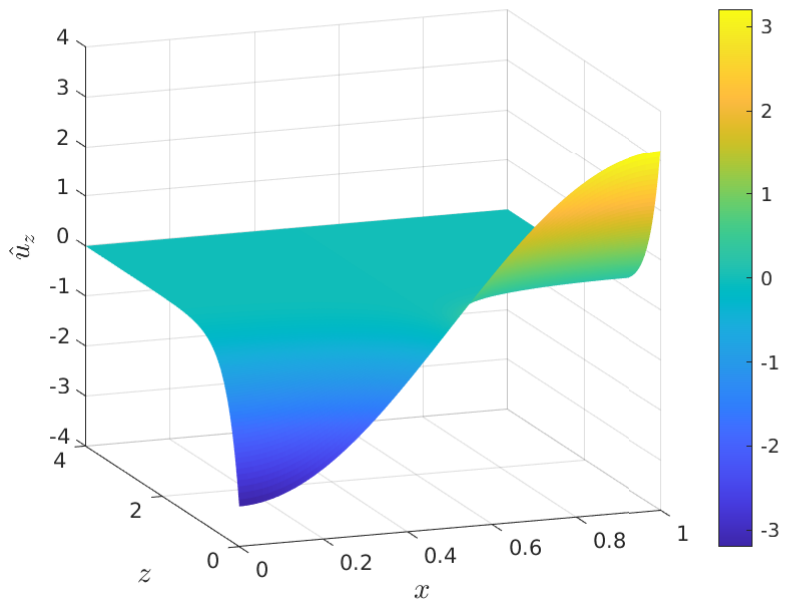}
         \caption{Computed values of $\hat u_z$ from the FEA solution.}
     \end{subfigure}
     \caption{\textcolor{cyan}{Comparison of the analytically and FEA computed functions $\hat u_x$ and $\hat u_z$ corresponding to the velocity field of a planar inviscid capillary wave with $n=1$. The FEA values were computed for $\textsf{Re}=710$}, and the analytical values follow as $\hat u=\nabla \hat\phi_n$ for $\hat \phi_n$ in \eqref{eq:ansatzexactsolution}.}
     \label{fig:velocity_cap}
\end{figure}

Finally, Fig. \ref{fig:velocity_cap} compares the profiles of the velocity components from \textcolor{cyan}{the FEA solution for a case \footnote{\textcolor{cyan}{This case was inspired by a domain filled with liquid aluminum in contact with argon at atmospheric conditions, which is discussed in more detail in Section \ref{subsec:cylinder}.}} with $\textsf{Re}=710$} with \textcolor{cyan} {its analytical inviscid counterpart}  $\hat {\bf u}_1= \nabla \hat \phi_1$, where $\hat\phi_n$ is given by \eqref{eq:ansatzexactsolution} for $n=1$. Note that for this example $\hat {\bf u}_1$ can  be chosen to have all real components. In contrast, the FEA solution is computed with a small value of $\textsf{Re}^{-1}$, so it is not necessarily possible to make all components of the computed velocity field real. However, the FEA solution can be multiplied by a complex scalar to make the real part of the components much larger than the imaginary part (in this case two orders of magnitude larger), and normalized so that the analytical and numerical solution have the same maximum value for some field, in this case, the absolute value of the real part of the velocity component in the $x-$direction. We show the real part of the resulting velocity components in the figure.

\subsubsection{Viscous Case}

The inviscid flow modes can be used to estimate the angular frequency and damping rate for the viscous flow modes  for small enough viscosity values. To do this, we will assume that the inviscid flow modes are good approximations to the viscous flow modes, and use them to solve for $\lambda$. Specifically, consider an incompressible mode $(\hat{\bf u},\hat p, \hat \xi)$  with complex frequency $\lambda$, whose meniscus surface displacement satisfies $\hat\xi=u_z/\lambda$, from \eqref{eq:2c}. Then, setting ${\bf v}=\overline{\hat {\bf u}}$ in  \eqref{eq:weaku}   implies that $\lambda$ must satisfy
\begin{equation}
\label{eq:evproblem}
    0=\lambda a(\hat{\bf u},\overline{\hat {\bf u}}) + c(\hat{\bf u},\overline{\hat{\bf u}}) + \frac{1}{\lambda}s(\hat u_z, \overline{\hat u_z}),
\end{equation}
where we used $b(\overline{\hat{\bf u}},p)=0$ because of the incompressibility of $\hat{\bf u}$. This same equation can alternatively be obtained directly from a mechanical energy balance. To estimate the damping rate, we solve \eqref{eq:evproblem} for $\lambda$ by replacing
\begin{align}
\label{vel}
    \hat {\bf u}= \nabla \hat \phi_n,
\end{align}
where $\hat \phi_n$ is given by \eqref{eq:ansatzexactsolution}. In the limit of infinite channel height ($H\to\infty$), we obtain
\begin{align}
\label{eq:lambda_viscous}
    \lambda_n = -2\frac{k_n^2}{\textsf{Re}} \pm i \sqrt{k_n^3\left(1 - 4\frac{k_n}{\textsf{Re}^2}\right)}.
\end{align}
Here, the (negative) real part corresponds to the damping rate, while the absolute value of the imaginary part corresponds to the angular frequency.
For weak damping ($\textsf{Re}\gg k_n^{1/2}$),  the damping rate  is $\eta_n=-2 k_n^2/\textsf{Re}$ and the angular frequency will approach $k_n^{3/2}$, which corresponds to $\omega_n$ for the inviscid case (see \eqref{eq:analyticalfrequency}).

\begin{figure}
\includegraphics[width=0.8\linewidth]{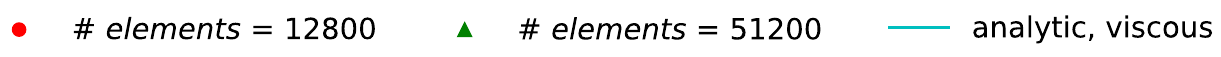}
     \centering
     \begin{subfigure}[b]{0.32\textwidth}
         \centering
         \includegraphics[width=\textwidth]{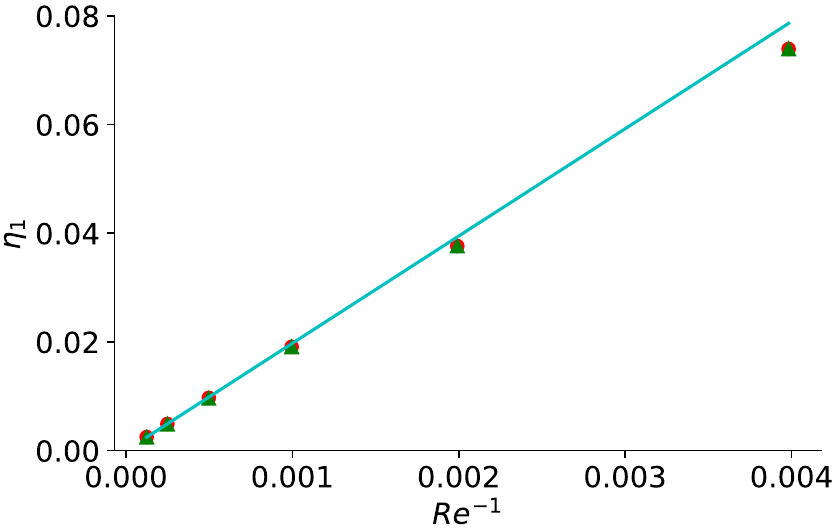}
         \caption{Mode $n=1$}
     \end{subfigure}
     \hfill
     \begin{subfigure}[b]{0.32\textwidth}
         \centering
         \includegraphics[width=\textwidth]{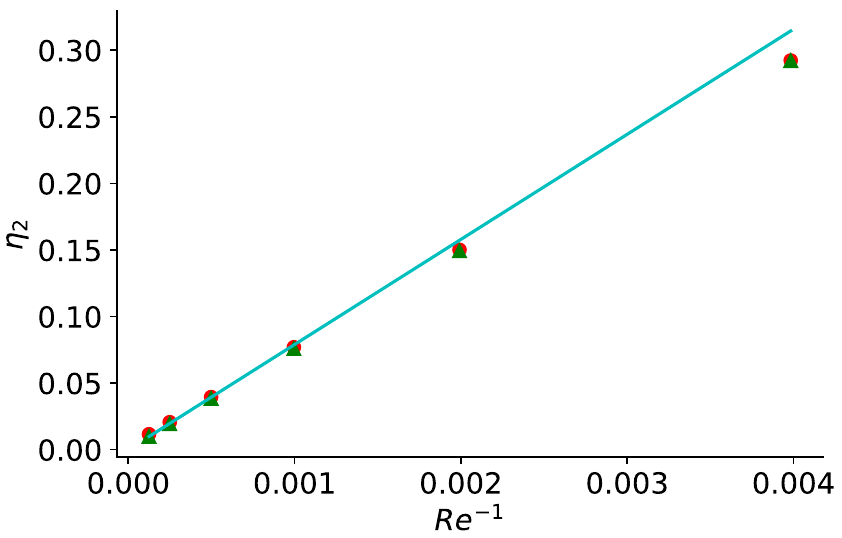}
         \caption{Mode $n=2$}
     \end{subfigure}
     \hfill
     \begin{subfigure}[b]{0.32\textwidth}
         \centering
         \includegraphics[width=\textwidth]{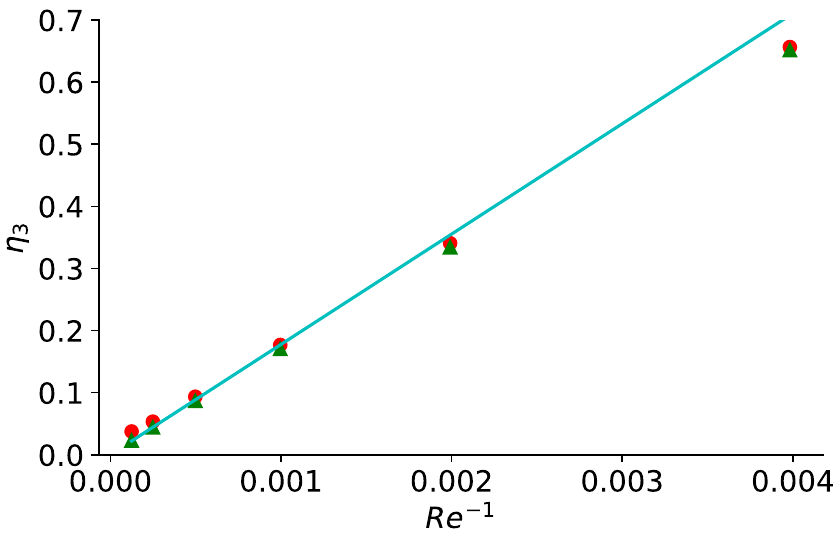}
         \caption{Mode $n=3$}
     \end{subfigure}
     \caption{Comparison of the damping rates obtained from the FEA solutions for the three lowest-angular-frequency modes of a planar viscous capillary wave with their counterparts computed via \eqref{eq:lambda_viscous} for (a) mode 1, (b) mode 2, and (c) mode 3. \textcolor{cyan}{As in Fig. \ref{fig:modes_frequencies}, the values computed from the FEA solutions are largely mesh-converged, as a comparison between the two levels of refinement suggests.}}    
     \label{fig:modes_damping_rates}
\end{figure}

Figure \ref{fig:modes_frequencies} compares the angular frequencies from \eqref{eq:lambda_viscous} with the estimates from the FEA solutions \textcolor{cyan} {with Reynolds numbers ranging from 251 to 8034} for $n=1,2,3$. Figure \ref{fig:modes_damping_rates} shows a similar comparison for the damping rates. For both quantities, we find a good agreement between the FEA estimates and the values obtained from \eqref{eq:lambda_viscous}, with deviations becoming larger as $\textsf{Re}^{-1}$ increases and the use of inviscid flow modes is no longer justified in deriving \eqref{eq:lambda_viscous}.

\subsection{Mesh Convergence on a Cylindrical Domain}
\label{subsec:cylinder}

Next, we compute the least-damped eigenmodes for capillary wave oscillations in a brimful cylinder, and evaluate the convergence of the computed eigenvalues and eigenvectors as the mesh size is decreased. Specifically, we consider an axisymmetric problem in the 2D domain, specified by $\breve{\Omega} = \big\{(r,z) ~\big|~ r\in[0,1) \text{ and }z\in (0,H)\}$, for $H=2.4$, with $\gamma_m=(0,1)\times\{0\}$, $\gamma_w=\{1\}\times(0,H)$, and $\gamma_t=(0,1)\times\{H\}$.  We set the Reynolds number to  ${\textsf{Re}=710}$, which was inspired by the case of $\breve\Omega$ having a dimensional length of $R=5\times 10^{-4}$ m (taken as the characteristic length) and being filled with liquid aluminum with density $\rho=2435\ \mathrm{kg/m^3}$ and dynamic viscosity $\nu=4.16\times 10^{-7}\ \mathrm{m^2/s}$, and the gas below $\gamma_m$ to be argon at atmospheric conditions such that the surface tension at $\gamma_m$ is $\sigma=0.85\ \mathrm{N/m}$.

\begin{figure}
     \centering
     \begin{subfigure}[b]{0.32\textwidth}
         \centering
         \includegraphics[width=\textwidth]{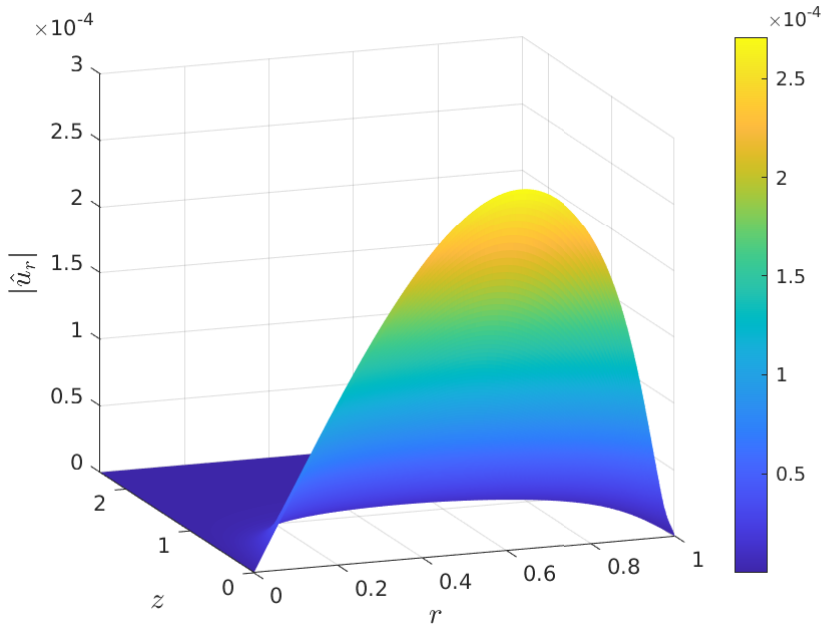}
         \caption{$|\hat u_r|$ for mode 1}
     \end{subfigure}
     \begin{subfigure}[b]{0.32\textwidth}
         \centering
         \includegraphics[width=\textwidth]{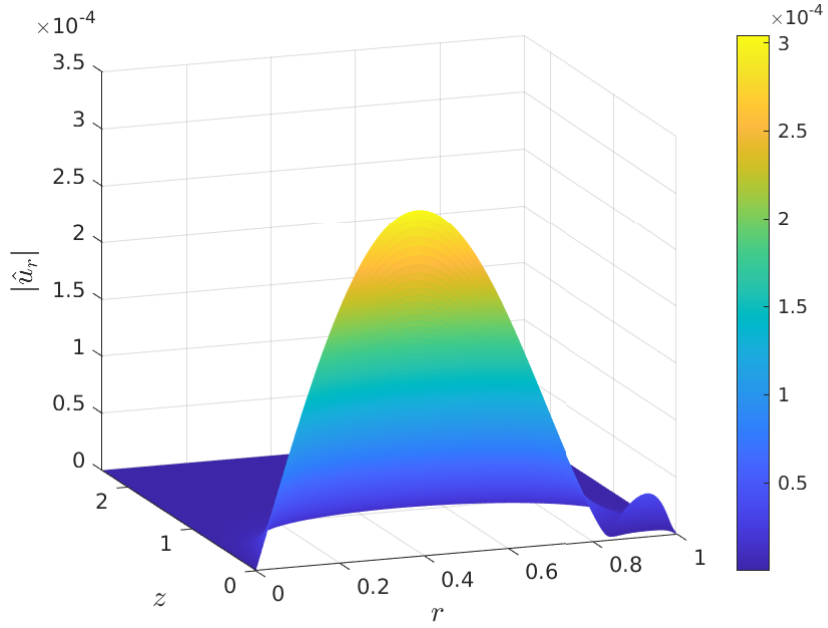}
         \caption{$|\hat u_r|$ for mode 2}
     \end{subfigure}
     \begin{subfigure}[b]{0.32\textwidth}
         \centering
         \includegraphics[width=\textwidth]{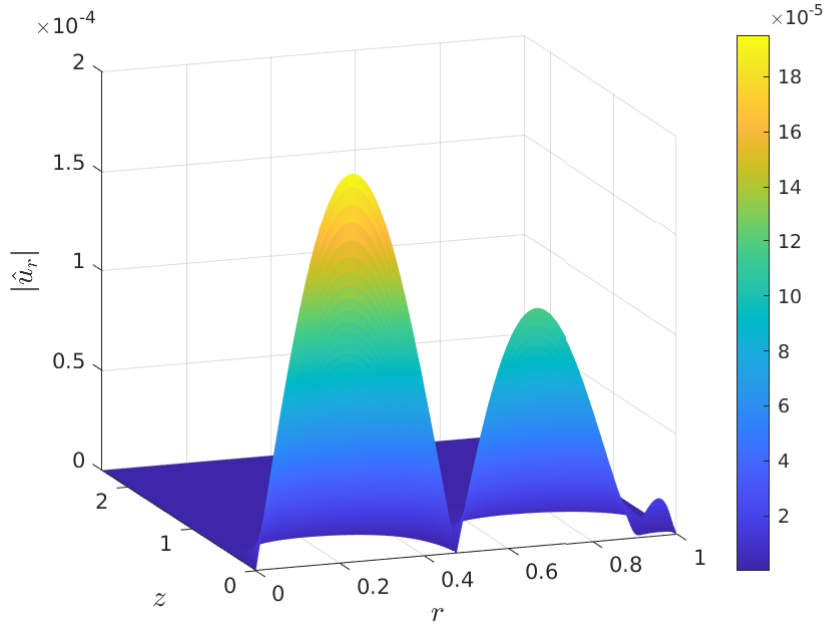}
         \caption{$|\hat u_r|$ for mode 3}
     \end{subfigure}
     \caption{Magnitude of $\hat u_r$ for the three least-damped modes in a cylindrical nozzle.}
     \label{fig:velocity_cyl}
\end{figure}

Figure \ref{fig:velocity_cyl} shows the computed magnitude of $\hat u_r$, or $(\hat u_r \overline{\hat u_r})^{1/2}$, for the three least-damped  modes. For notational convenience, we will refer to them below as modes 1, 2 and 3, in order of increasing damping rate.

\begin{figure}
    \centering
    \includegraphics[width=0.8\textwidth]{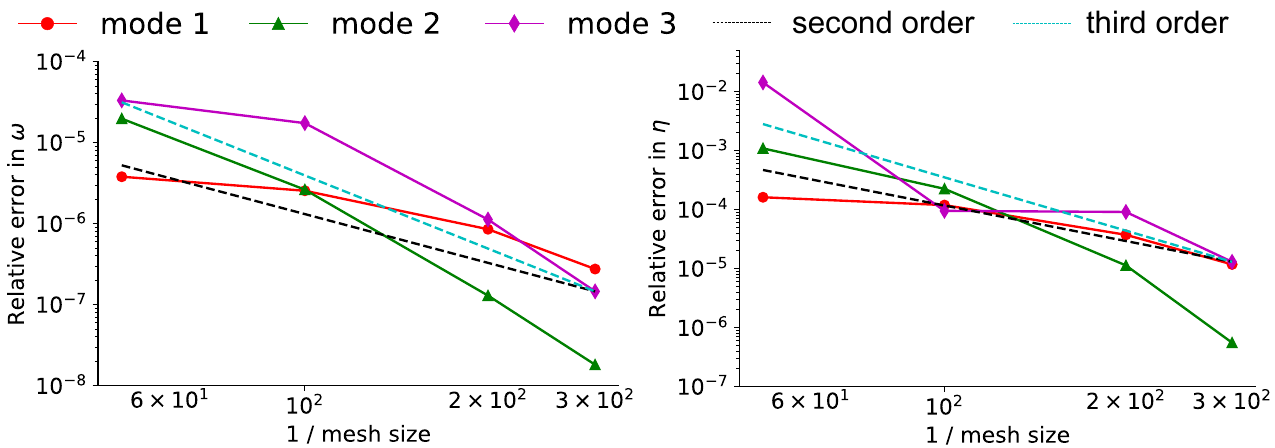}
    \caption{Convergence curves for the computed angular frequency $\omega$ (left) and damping rate $\eta$ (right) for the three least-damped modes in a cylindrical domain.}
    \label{fig:freq_damp_cyl}
\end{figure}

To evaluate the convergence of the method in this example, we consider a sequence of uniform meshes with decreasing element size, and examine the convergence of the eigenvalues (angular frequencies and damping rates) and associated eigenvectors (velocity, pressure, and meniscus surface displacement) over this sequence. Figure \ref{fig:freq_damp_cyl} shows the relative errors $(\omega-\omega_\text{finest})/\omega_\text{finest}$ and $(\eta-\eta_\text{finest})/\eta_\text{finest}$ for the three least-damped modes as a function of mesh resolution, indicated by the total number of finite elements in the computational domain. Here $\omega_\text{finest}$ and $\eta_\text{finest}$ are the values computed with the mesh that has the highest number of elements. The convergence of these differences is strong evidence that that the computed values converge for all three modes. For both $\omega$ and $\eta$, the convergence rate is between second and third order for mode 1, and faster than third order for modes 2 and 3.

Next, we investigate mesh convergence for the eigenvectors. To easily compare eigenvectors from different meshes, we define a regular Cartesian mesh $\mathcal T_\text{cart}$  that covers the entire domain of the problem, and whose mesh size is finer than the finest finite element mesh we use. Part of the boundary of this Cartesian mesh exactly meshes $\gamma_m$. Given a function $f\colon \breve\Omega\to\mathbb C$ or $f\colon\gamma_m\to \mathbb C$, we  define the vector of interpolated values ${\cal I}f$ whose $i$-th component is the value of $f$ at the $i$-th vertex of $\mathcal T_\text{cart}$. If $f$ is not defined on the $i$-th vertex, the $i$-th component of $\mathcal I f$ is set to zero. As a result, the $i$-th component of $ \mathcal I \hat \xi_h$ is set to zero if the $i$-th vertex is not on $\gamma_m$. Therefore, given a computed  eigenvector $(\hat{\bf u}_h,\hat p_h, \hat \xi_h)$ over a finite element mesh, we compute the vector of interpolated values $X_h=({\cal I}(\hat u_r)_h,{\cal I}(\hat u_z)_h, {\cal I}\hat p_h,{\cal I}\hat \xi_h)$, and normalize it so that $X_h^{\sf T}\overline{X}_h=1$. For simplicity of notation, for any such interpolated eigenvector $X_h$, we denote the components associated to each one of the fields as $X_h^r={\cal I}(\hat u_r)_h$, $X_h^z={\cal I}(\hat u_z)_h$, $X_h^p={\cal I}\hat p_h$, and $X_h^\xi={\cal I}\hat \xi_h$. Each one of these vectors has a length equal to the number of vertices in the Cartesian mesh. Finally, we can compute the Euclidean norm of any of these vectors as, for example, $|X_h^r|=((X_h^r)^{\sf T} \overline{X_h^r})^{1/2}$, and so on.

Complex eigenvectors are defined up to a complex scalar, i.e., if $X$ is an  eigenvector, so is $\beta X$ for any $\beta\in \mathbb C$. To compare two interpolated eigenvectors $X_1$ and $X_2$ for convergence studies, we find $\beta$  as
\begin{equation}
    \label{eq:adef}    
    \beta(X_1,X_2) = \argmin_{\beta\in \mathbb C,|\beta|=1} |X_1-\beta X_2|^2 =\overline{X_2^{\sf T}\overline{X}_1}/|X_2^{\sf T}\overline{X}_1|,
\end{equation}
and compute the difference between the two as 
\begin{displaymath}
    \Delta X(X_1,X_2) = X_1-\beta(X_1,X_2) X_2.
\end{displaymath}
The computation of $\beta(X_1,X_2)$ is shown in \ref{sec:adef}.

\begin{figure}
 \centering
 \includegraphics[width=0.6\textwidth]{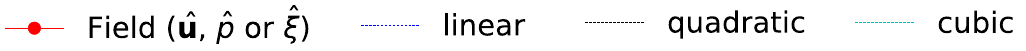}\\
    \begin{subfigure}{0.32 \textwidth}
        \includegraphics[width=\textwidth]{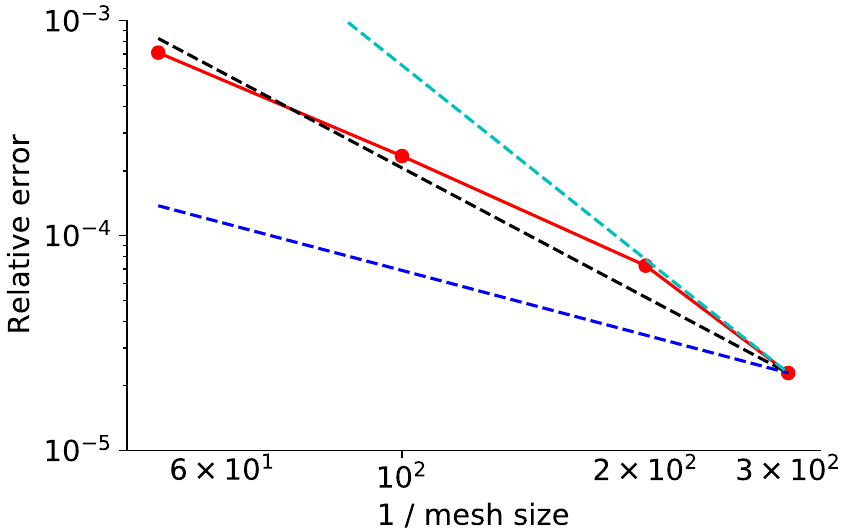}
        \subcaption{$e_u^h$ for mode 1}
    \end{subfigure}
    \begin{subfigure}{0.32 \textwidth}
        \includegraphics[width=\textwidth]{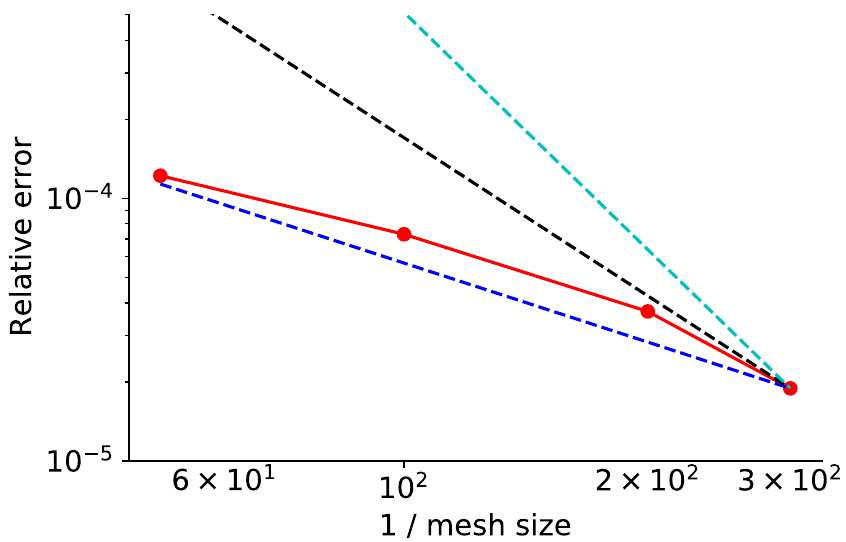}
        \subcaption{$e_p^h$ for mode 1}
    \end{subfigure}
    \begin{subfigure}{0.32 \textwidth}
        \includegraphics[width=\textwidth]{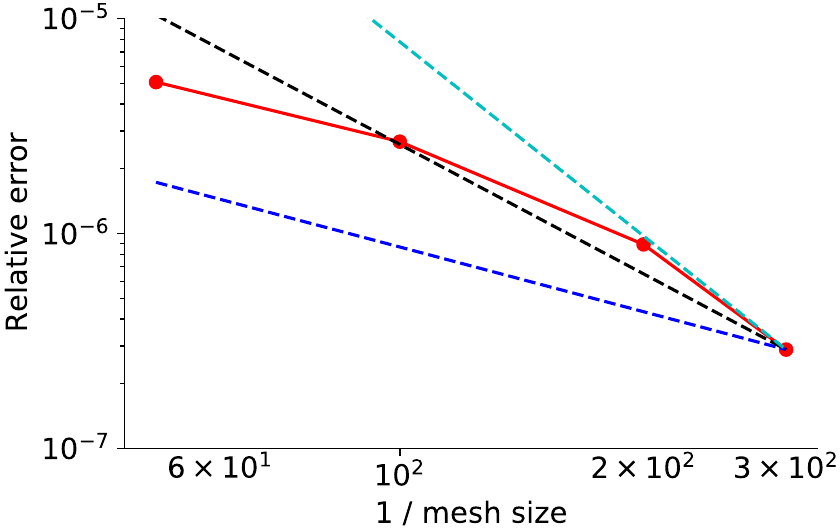}
        \subcaption{$e_\xi^h$ for mode 1}
    \end{subfigure}
    \begin{subfigure}{0.32 \textwidth}
        \includegraphics[width=\textwidth]{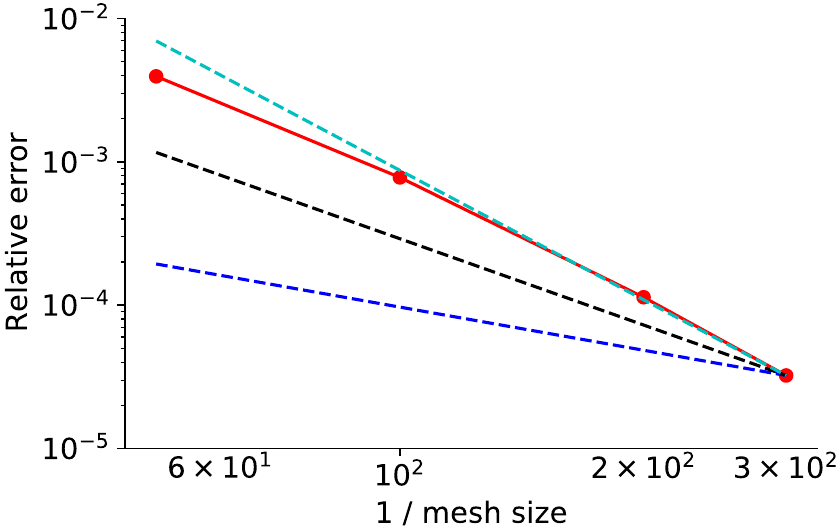}
        \subcaption{$e_u^h$ for mode 2}
    \end{subfigure}
    \begin{subfigure}{0.32 \textwidth}
        \includegraphics[width=\textwidth]{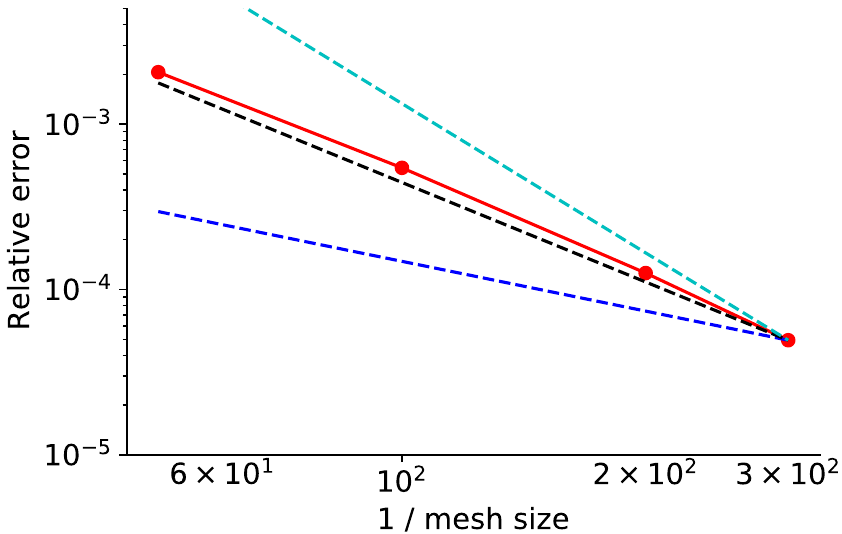}
        \subcaption{$e_p^h$ for mode 2}
    \end{subfigure}
    \begin{subfigure}{0.32 \textwidth}
        \includegraphics[width=\textwidth]{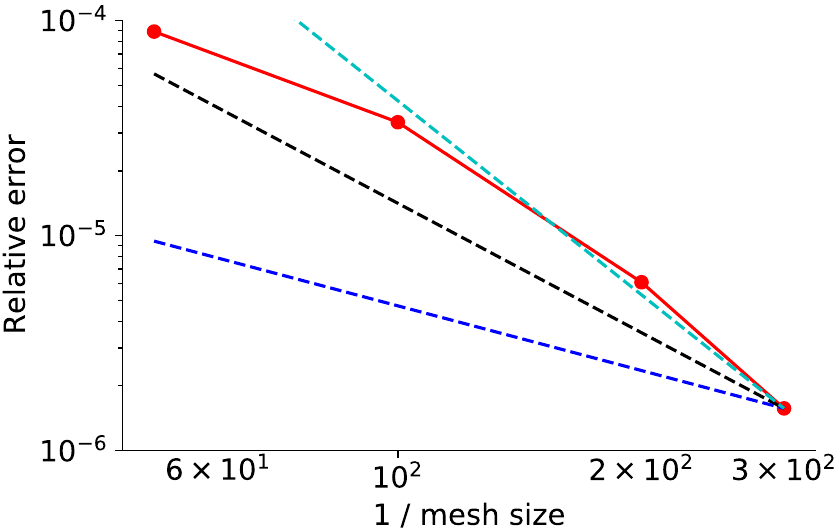}
        \subcaption{$e_\xi^h$ for mode 2}
    \end{subfigure}
    \begin{subfigure}{0.32 \textwidth}
        \includegraphics[width=\textwidth]{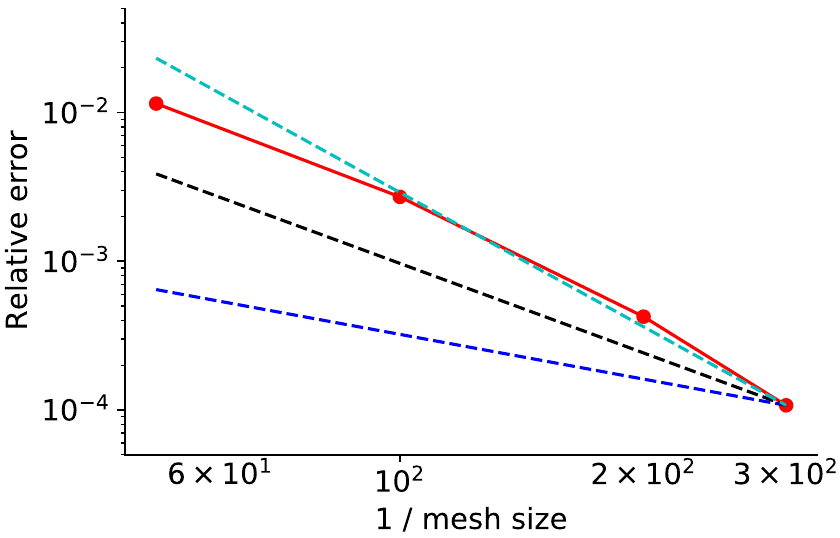}
        \subcaption{$e_u^h$ for mode 3}
    \end{subfigure}
    \begin{subfigure}{0.32 \textwidth}
        \includegraphics[width=\textwidth]{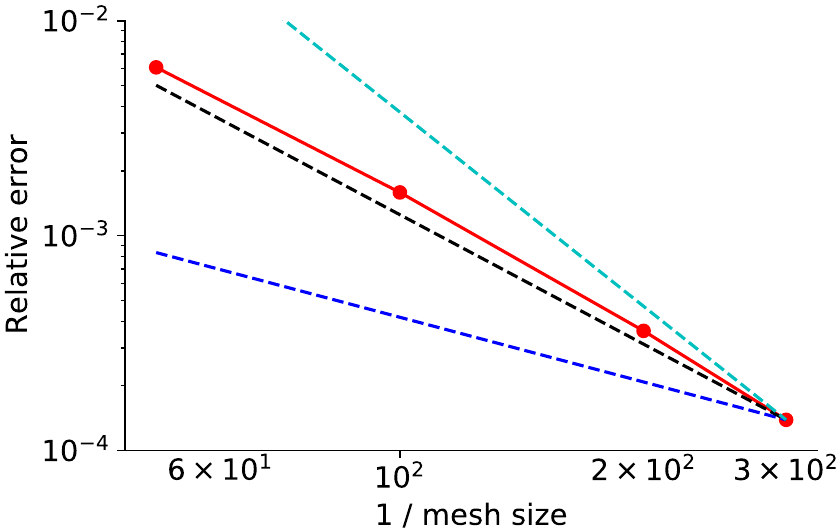}
        \subcaption{$e_p^h$ for mode 3}
    \end{subfigure}
    \begin{subfigure}{0.32 \textwidth}
        \includegraphics[width=\textwidth]{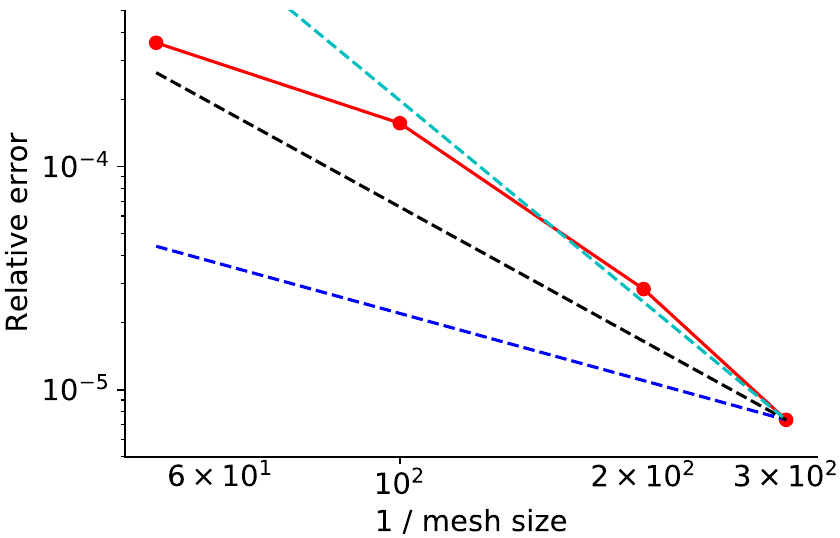}
        \subcaption{$e_\xi^h$ for mode 3}
    \end{subfigure}
    \caption{Convergence curves for the velocity $e_u^h$ (left), pressure $e_p^h$ (middle), and meniscus surface displacement $e_\xi^h$ (right) for modes 1 (top), 2 (middle), and 3 (bottom) in a cylindrical domain.}
    \label{fig:eigenvecs_cyl}
\end{figure}

As a proxy for the exact solution, we compute errors with respect to the interpolation of the eigenvector obtained on the finest finite element mesh we used, denoted $X_{f}$. We compute the relative errors
\begin{align*}
    e_u^h&=\left[\frac{|\Delta X(X_h^r,X_f^r)|^2+|\Delta X(X_h^z,X_f^z)|^2}{|X_f^r|^2+|X_f^z|^2}\right]^{1/2}, \\
    e_p^h & = \frac{|\Delta X(X_h^p, X_f^p)|}{|X_f^p|}, \\
    e_\xi^h&= \frac{|\Delta X(X_h^\xi, X_f^\xi)|}{|X_f^\xi|},
\end{align*}
for modes 1, 2 and 3 in Fig. \ref{fig:eigenvecs_cyl} as a function of the mesh size. These are approximations of the relative $L^2-$errors of each one the fields. Third-order convergence is observed for the velocities and the meniscus displacement, while second-order convergence is observed for the pressure field.

\subsection{Mesh Convergence on an Arbitrary Axisymmetric Domain}
\label{subsec:arbitrary_shape}

\begin{figure}[htb]
    \centering
    \includegraphics[width=0.7\linewidth]{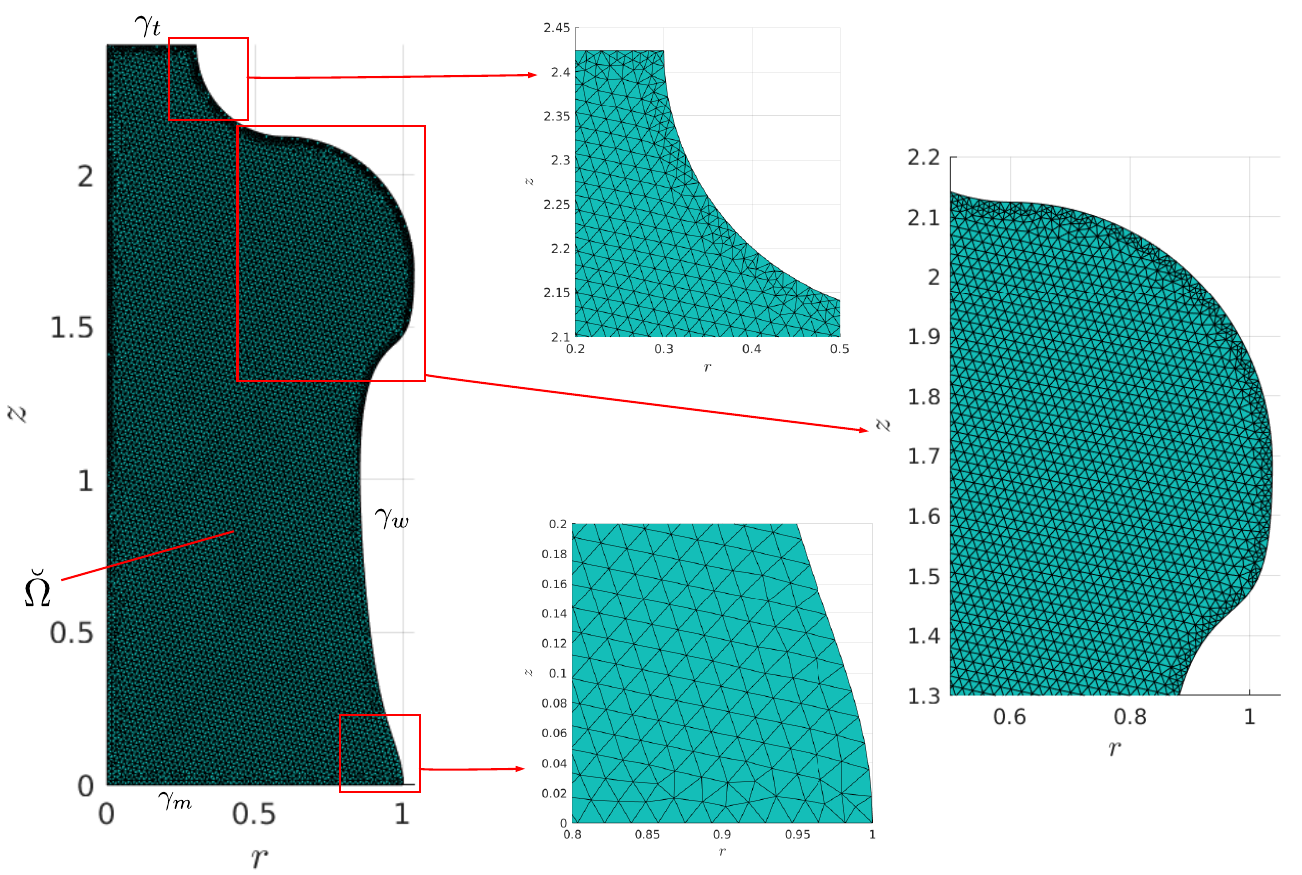}
    \caption{Domain $\breve\Omega$ used to illustrate the performance of the method in arbitrary axisymmetric geometries.}
    \label{fig:axisymm_domain}
\end{figure}

\begin{figure}
     \centering
     \begin{subfigure}[b]{0.32\textwidth}
         \centering
         \includegraphics[width=\textwidth]{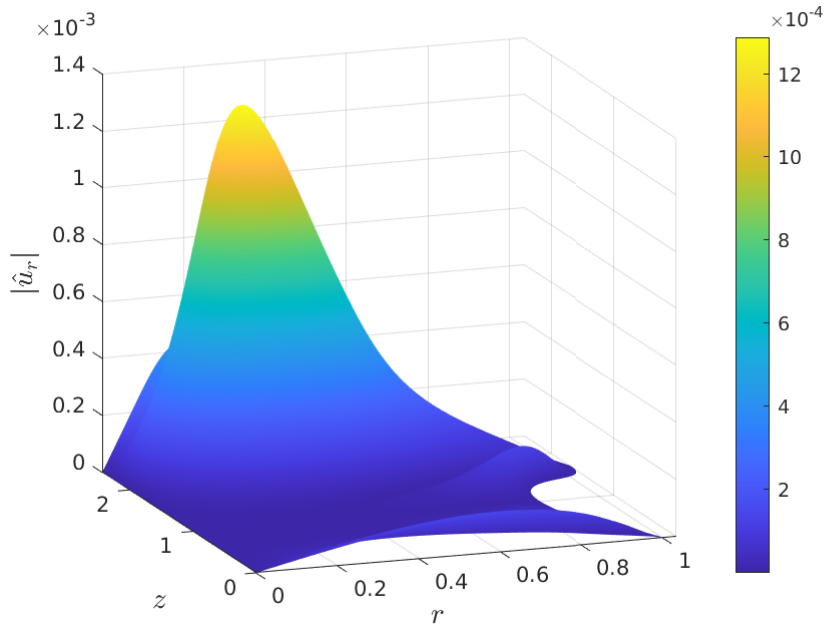}
         \caption{$|\hat u_r|$ for mode 1}
     \end{subfigure}
     \begin{subfigure}[b]{0.32\textwidth}
         \centering
         \includegraphics[width=\textwidth]{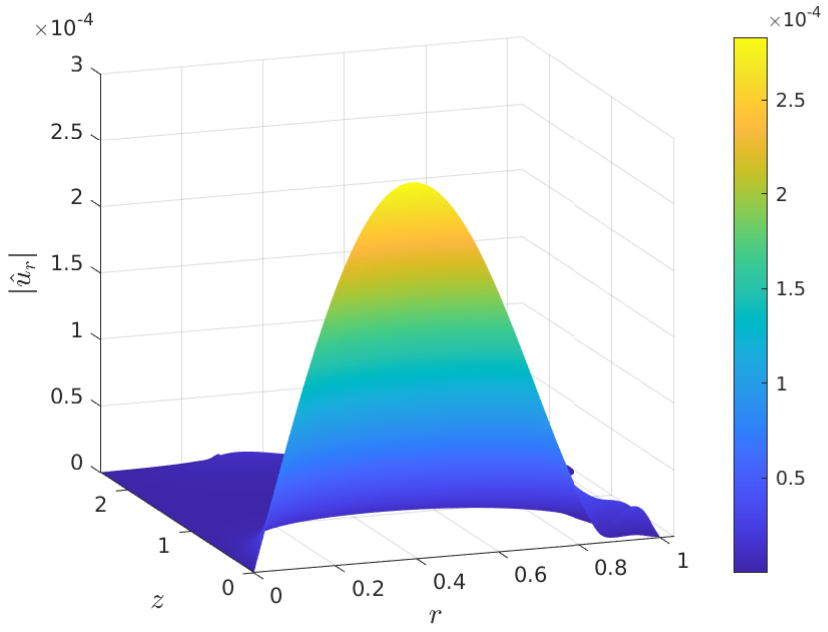}
         \caption{$|\hat u_r|$ for mode 2}
     \end{subfigure}
     \begin{subfigure}[b]{0.32\textwidth}
         \centering
         \includegraphics[width=\textwidth]{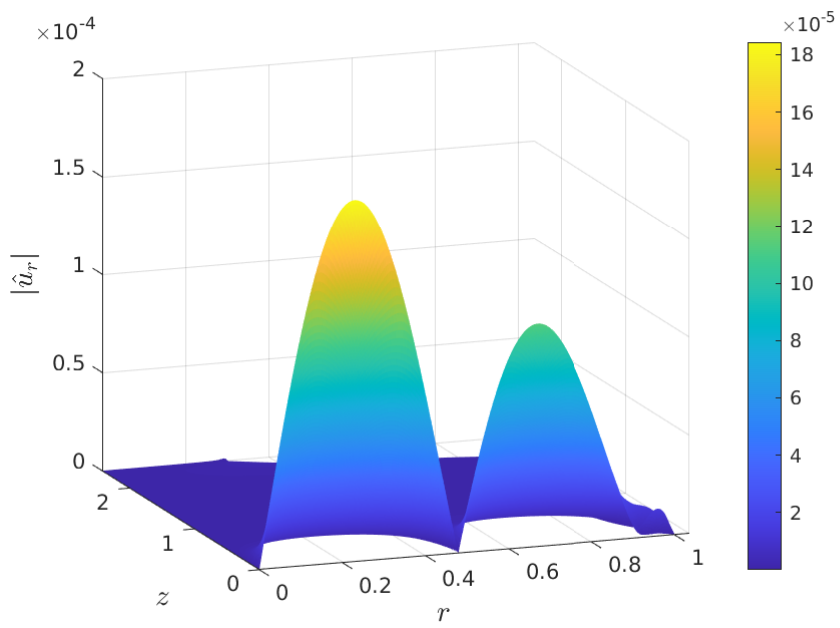}
         \caption{$|\hat u_r|$ for mode 3}
     \end{subfigure}
     \caption{Magnitude of $\hat u_r$ for the three least-damped modes in the nozzle shown in Fig. \ref{fig:axisymm_domain}.}
     \label{fig:velocity_arb}
\end{figure}

The last example of the paper demonstrates the use of the algorithm in axisymmetric domains with more complex geometry. In this example we compute the least-damped modes for the domain $\breve\Omega$ shown in Fig. \ref{fig:axisymm_domain}. We constructed the geometry and mesh in \gmsh by modifying the ``\texttt{chess pawn.geo}'' file obtained from \cite{NTNU2013}. Once again, we set the Reynolds number to ${\textsf{Re}=710}$ as in \S \ref{subsec:cylinder}.
Figure \ref{fig:velocity_arb} shows the computed magnitude of $\hat u_r$, or $(\hat u_r \overline{\hat u_r})^{1/2}$, for the three least-damped modes. 

\begin{figure}
     \centering
     \includegraphics[width=0.8\textwidth]{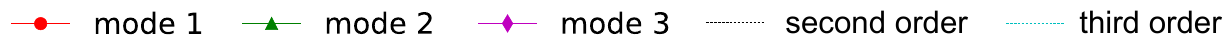}
     \begin{subfigure}[b]{0.4\textwidth}
         \centering
         \includegraphics[width=\textwidth]{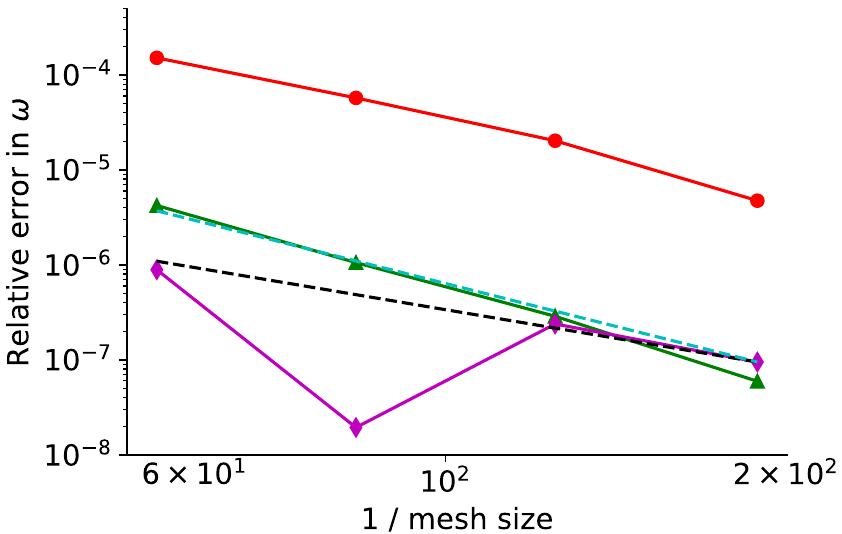}
     \end{subfigure}
     \begin{subfigure}[b]{0.4\textwidth}
         \centering
         \includegraphics[width=\textwidth]{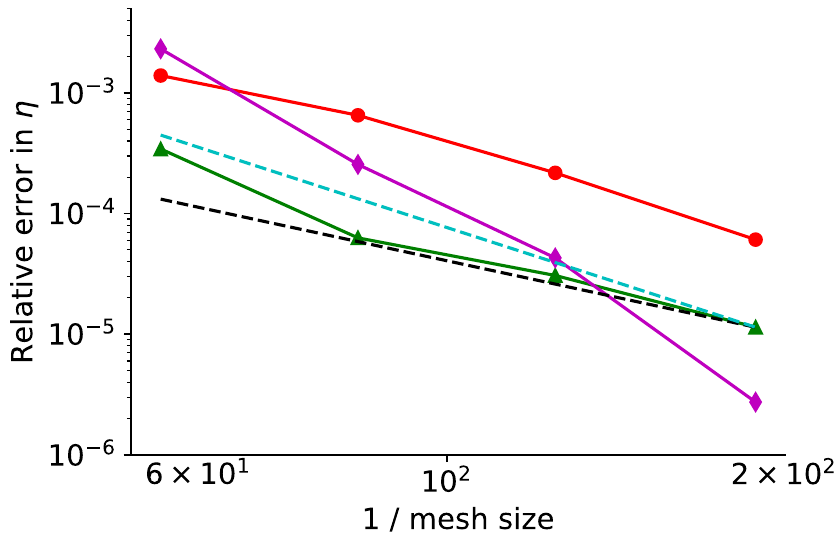}
     \end{subfigure}
     \caption{Convergence curves for the computed angular frequency $\omega$ (left) and damping rate $\eta$ (right) for the three least-damped modes in the domain of Figure \ref{fig:axisymm_domain}.}
     \label{fig:freq_damp_arb}
\end{figure}

To evaluate mesh convergence, once again we consider a sequence of meshes with decreasing element size\footnote{Unlike for the cylindrical domain in \S \ref{subsec:cylinder} due to the irregular shape of $\gamma_w$ the mesh will be mostly, but not entirely, uniform as it needs to conform to $\gamma_w$.}, and examine the convergence of the eigenvalues and associated eigenvectors of these modes over this sequence. Figure \ref{fig:freq_damp_arb} shows the relative errors $(\omega-\omega_\text{finest})/\omega_\text{finest}$ and $(\eta-\eta_\text{finest})/\eta_\text{finest}$ for the three least-damped modes as a function of mesh resolution, indicated by the inverse of the grid size parameter used in \gmsh, where $\omega_\text{finest}$ and $\eta_\text{finest}$ are defined as before. We can see that the method is mesh-convergent in both the angular frequency and damping rate for all three modes. 
The convergence rate is between 2 and 3 in all cases, with the damping rate converging at a faster rate for mode 3.

\begin{figure}
 \centering
 \includegraphics[width=0.6\textwidth]{legend_fields.pdf}\\
    \begin{subfigure}{0.32 \textwidth}
        \includegraphics[width=\textwidth]{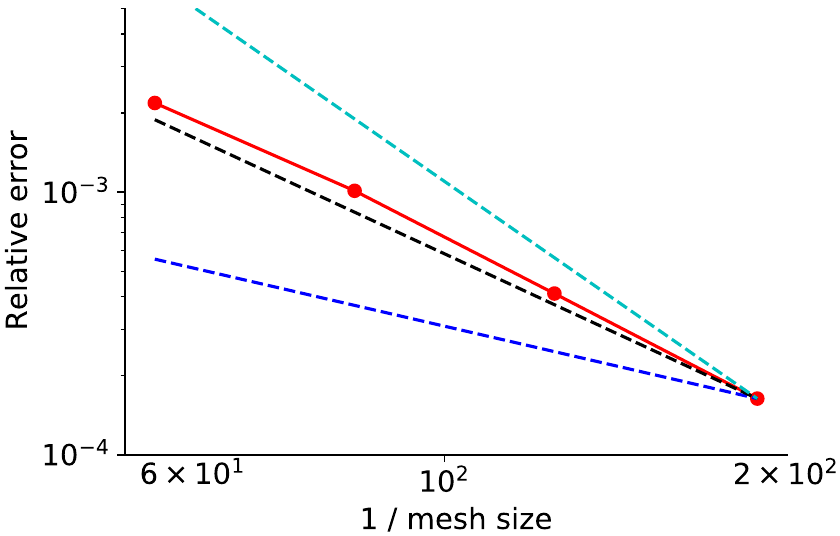}
        \subcaption{$e_u^h$ for mode 1}
    \end{subfigure}
    \begin{subfigure}{0.32 \textwidth}
        \includegraphics[width=\textwidth]{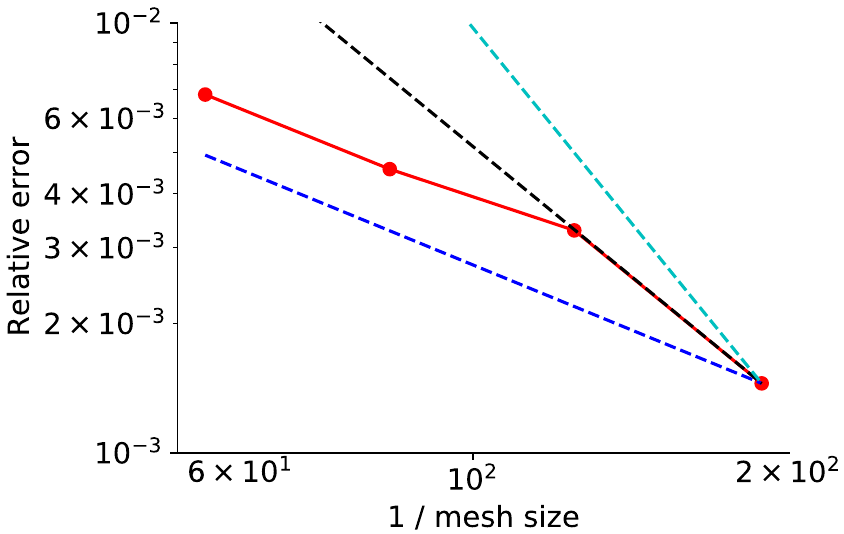}
        \subcaption{$e_p^h$ for mode 1}
    \end{subfigure}
    \begin{subfigure}{0.32 \textwidth}
        \includegraphics[width=\textwidth]{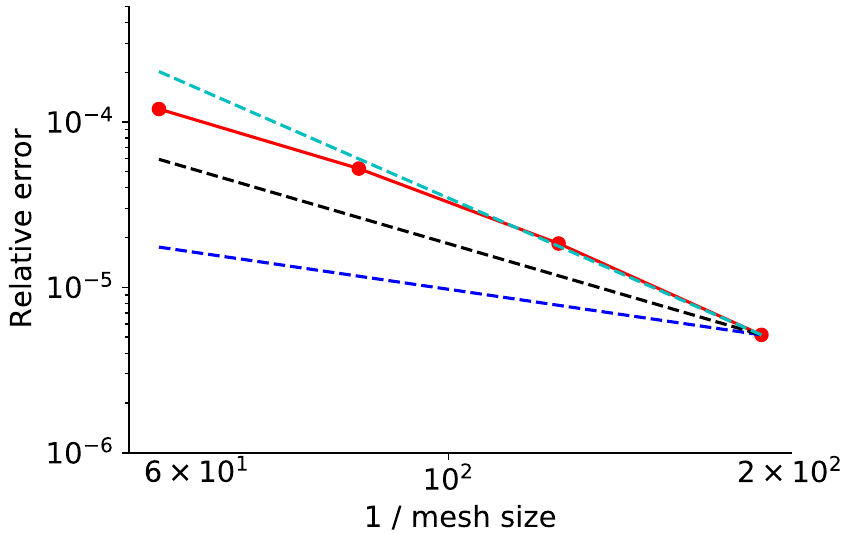}
        \subcaption{$e_\xi^h$ for mode 1}
    \end{subfigure}
    \begin{subfigure}{0.32 \textwidth}
        \includegraphics[width=\textwidth]{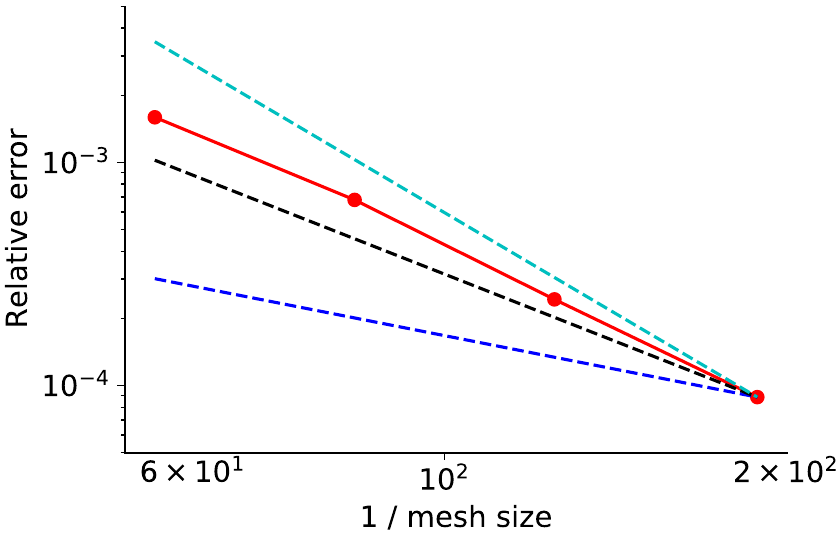}
        \subcaption{$e_u^h$ for mode 2}
    \end{subfigure}
    \begin{subfigure}{0.32 \textwidth}
        \includegraphics[width=\textwidth]{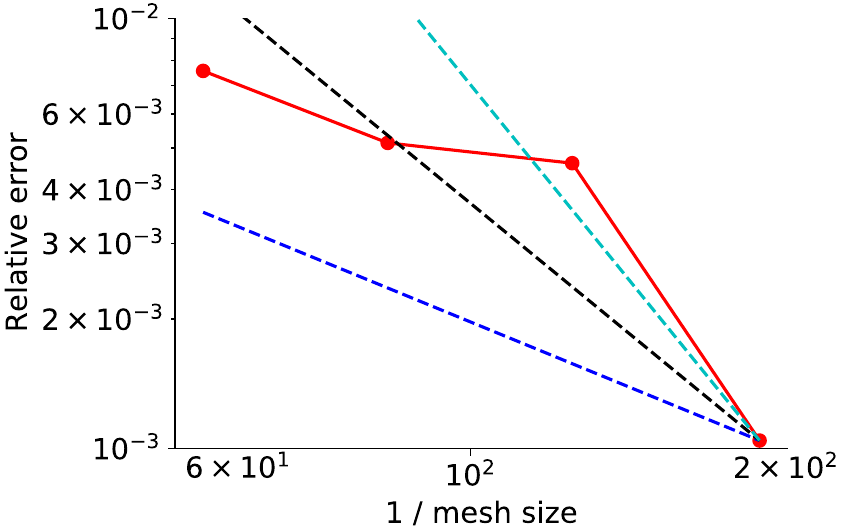}
        \subcaption{$e_p^h$ for mode 2}
    \end{subfigure}
    \begin{subfigure}{0.32 \textwidth}
        \includegraphics[width=\textwidth]{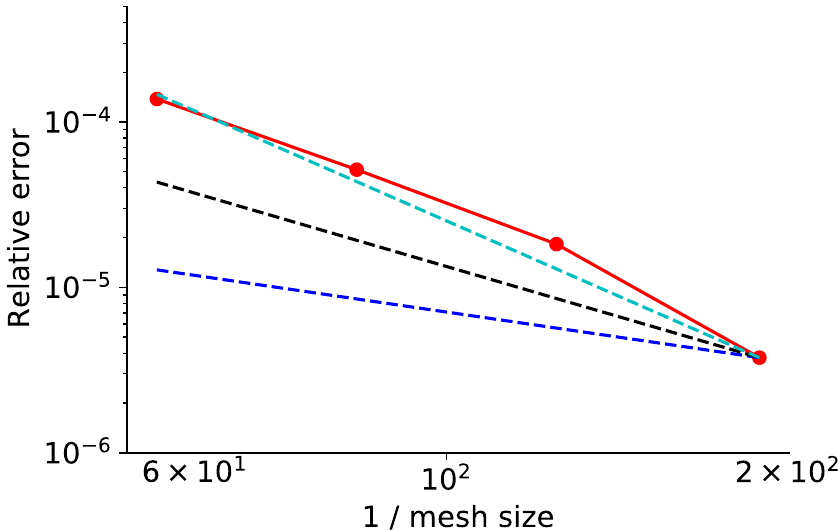}
        \subcaption{$e_\xi^h$ for mode 2}
    \end{subfigure}
    \begin{subfigure}{0.32 \textwidth}
        \includegraphics[width=\textwidth]{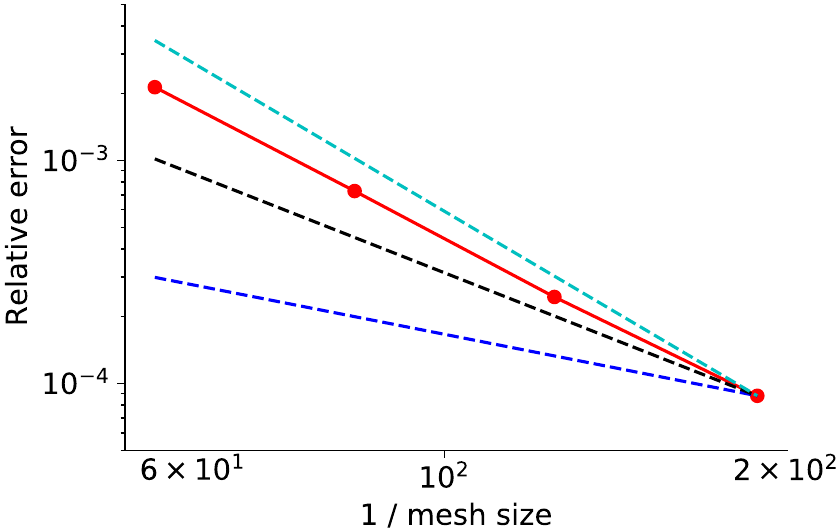}
        \subcaption{$e_u^h$ for mode 3}
    \end{subfigure}
    \begin{subfigure}{0.32 \textwidth}
        \includegraphics[width=\textwidth]{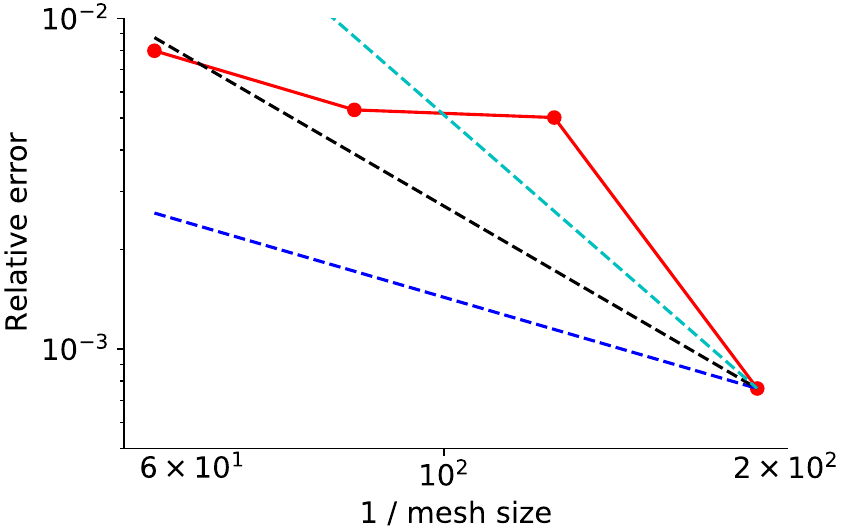}
        \subcaption{$e_p^h$ for mode 3}
    \end{subfigure}
    \begin{subfigure}{0.32 \textwidth}
        \includegraphics[width=\textwidth]{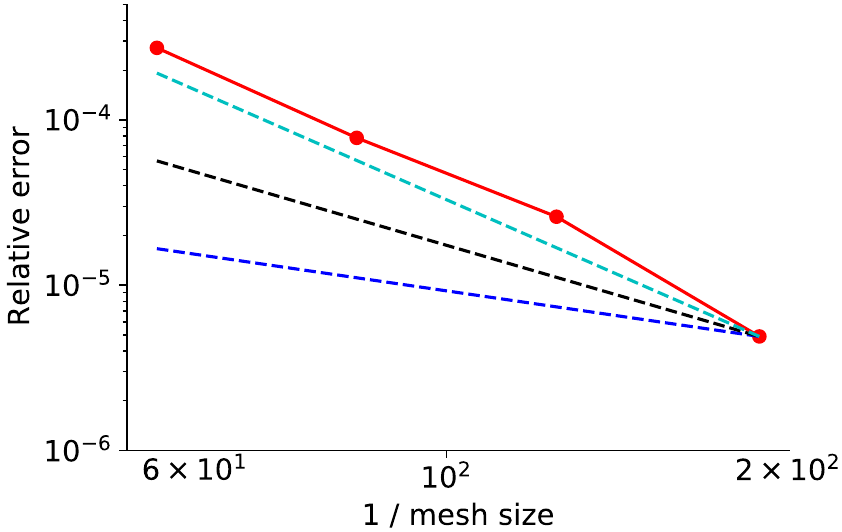}
        \subcaption{$e_\xi^h$ for mode 3}
    \end{subfigure}
    \caption{Convergence curves for the velocity $e_u^h$ (left)), pressure $e_p^h$ (middle), and meniscus surface displacement $e_\xi^h$ (right)n) for modes 1 (top), 2 (middle), and 3 (bottom) in the domain of Fig. \ref{fig:axisymm_domain}.}
    \label{fig:eigenvecs_arb}
\end{figure}

We observe similar results for the eigenvectors, after computing relative errors  as described in \S \ref{subsec:cylinder}. Figure \ref{fig:eigenvecs_arb} shows the relative errors for velocity, pressure and meniscus surface deformation. Compared to the cylindrical domain in \S \ref{subsec:cylinder}, we observe a similar third order (or higher) convergence for the meniscus displacement, while the velocities now display a convergence rate between second and third order. It is not possible to state the order of convergence of the pressure field in this example, but it is definitively second-order or higher.

%% file: summary_outlook.tex
\section{Summary \& Outlook}
\label{sec:conclusions}

We introduced a method to compute the late-time oscillation modes of a fluid with a free-surface in which viscosity is small and surface tension is high, a class of problems inspired by fluids in the nozzles of drop-on-demand (DoD) microfluidic devices. The approach is based on transforming the time-dependent problem into the computation of an eigensystem, in which the modes of oscillations are eigenmodes and their complex frequencies (which contains both their angular frequencies and damping rates) are the eigenvalues, as a way to circumvent the computational expense of computing \emph{converged} time-dependent evolutions through CFD. To this end, we constructed a finite element method, and showed that if some mild conditions are satisfied by the finite element spaces for the velocity, pressure, and meniscus displacement fields, then the discretization is guaranteed to return only eigenvalues with a negative real part. One set of spaces that satisfies these conditions are Taylor-Hood spaces for the velocity and pressure fields, and continuous piecewise quadratic elements for the meniscus displacements.

Through a suite of numerical tests, we verified the accuracy of the finite element analysis (FEA) solutions both in terms of the computed oscillation modes  and the corresponding angular frequencies and damping rates:
\begin{enumerate}
    \item The FEA solutions correctly approximate the analytical modal angular frequencies and damping rates of a viscous capillary wave in the limit of small viscosity.
    \item In a cylindrical domain, the FEA solution achieves at least second-order convergence in angular frequency and damping rate. The convergence rate of the eigenvectors coincides with that expected from the approximation properties of Taylor-Hood elements: second order in the pressure field and third order in the velocities and meniscus surface displacement.
    \item For a more generic axisymmetric domain, the FEA solution achieves at least second-order convergence in angular frequency and damping rate of the three least-damped modes. For the oscillation modes, the convergence rate is at least third order in the meniscus surface displacement, but between second and third order in the velocity field. The observed pressure field convergence rate is mixed, but at least second order or higher. 
\end{enumerate}

In future work, the current framework can be extended to handle arbitrary, non-axisymmetric nozzle shapes, and incorporate the effects of gravity by allowing the equilibrium meniscus shape to be curved. Moreover, the linearized solver here can be integrated into a constrained optimization loop to speed up nozzle prototyping, where the constraints may come from droplet specifications (e.g., user-defined mass or velocity), and minimum liquid capacity inside the nozzle, among others. Such a solver would facilitate the optimal design of nozzles in, for example, DoD printing devices for additive manufacturing.

\section*{Acknowledgments}

This research was developed with funding from the Xerox Corporation and from SRI International. The views, opinions, and/or findings expressed are those of the authors and should not be interpreted as representing the official views or policies of the Xerox Corporation or SRI International. Professor Adrian J. Lew's contributions to this publication was as a paid consultant and was not part of his Stanford University duties or responsibilities. The authors would like to thank Prof. Jose E. Roman for his helpful comments regarding the use of the \slepc eigensolver. Furthermore, they would like to thank Prof. John Burkardt for his advice on using and modifying routines from his library of open-source finite element \matlab codes.

%% file: appendix.tex
\section{Appendix}
\subsection{Preliminaries}

For ${\bf u}\in \mathcal V$ and $\xi\in \Xi$, it holds that
\begin{subequations}
\begin{align}
    c({\bf u},\overline{\bf u})&=0 \Longleftrightarrow {\bf u}={\bf 0}  \label{eq:prop1} \\
    s(\xi,\overline\xi)&=0 \Longleftrightarrow \xi={ 0},\label{eq:prop2}
\end{align}
where the bilinear forms $c$ and $s$ were defined in \eqref{eq:bilinearforms}. To see that \eqref{eq:prop1} holds, notice that $c({\bf u},\overline{\bf u})=0$ implies that $\nabla {\bf u}+(\nabla {\bf u})^{\sf T}=0$. Since ${\bf u}=0$ on $\Gamma_w$ from \eqref{eq:defnu} and $\Omega$ is connected, Korn's inequality \cite{ciarlet2021mathematical} helps us conclude that $\nabla {\bf u}=0$, and hence that ${\bf u}=0$. The converse is trivial.

To see that \eqref{eq:prop2} holds, notice that $s(\xi,\overline{\xi})=0$ implies that $\nabla_S\xi =0$ on $\Gamma_m$, or that $\xi$ is constant on $\Gamma_m$. Because $\xi=0$ on $\overline{\Gamma_m} \cap \overline{\Gamma_w}$, we conclude that $\xi=0$. Again, the converse is trivial.

Finally, since $c({\bf u},\overline{\bf u})\ge 0$, \eqref{eq:prop1} implies that 
\begin{equation}
    \label{eq:prop3}
    c({\bf u},\overline{\bf u})>{\bf 0} \Longleftrightarrow {\bf u}\neq{\bf 0}.
\end{equation}
\end{subequations}

\subsection{Derivation of the Weak Form}
\label{app:weakform}

Let $\hat\Sigma=-\hat p {\bf I}+\frac{1}{\text{Re}} (\nabla \hat{\bf u}+(\nabla \hat{\bf u})^{\sf T})$ denote the stress tensor, so that \eqref{eq:2}, \eqref{eq:3}, and \eqref{eq:5} become
\begin{subequations}
\begin{align}
\lambda \hat{\bf u} &= \nabla \cdot\hat{\Sigma} &&
\text{ in }\Omega,\label{eq:basiceq}\\
\hat \Sigma\cdot{\bf e}_z & = - \Delta_S\hat\xi {\bf e}_z && \text{ on }\Gamma_m,\label{eq:bc1instress}\\
\hat \Sigma\cdot{\bf e}_z & = 0 && \text{ on }\Gamma_t.\label{eq:bc2instress}
\end{align}
\end{subequations}
To obtain \eqref{eq:weak1}, we compute the inner product of \eqref{eq:basiceq} with any ${\bf v}\in \mathcal V$ and integrate over  $\Omega$, to get
\begin{align}
    0 & = \int_{\Omega}  \lambda \hat{\bf u} \cdot {\bf v} -(\nabla \cdot \hat \Sigma)\cdot{\bf v}\; d\Omega\nonumber \\
    & = \int_{ \Gamma_m} {\bf v}\cdot \hat \Sigma\cdot (-{\bf n}) \;dS  + \int_{\Omega}  \lambda \hat{\bf u} \cdot {\bf v} + \hat \Sigma \colon\nabla {\bf v} \;d\Omega\label{eq:partialresult}\\
    & = \int_{\Gamma_m} - (\Delta_S\hat  \xi) ~ v_z\; dS + \int_{\Omega}  \lambda \hat{\bf u} \cdot {\bf v} + \hat \Sigma \colon\nabla {\bf v} \;d\Omega\nonumber\\
    & = \int_{\Gamma_m}  \nabla_S\hat \xi \cdot\nabla_S v_z\; dS
    +  \int_{\Omega} \lambda  \hat{\bf u} \cdot {\bf v}+ \hat \Sigma\colon \nabla{\bf v}\;d\Omega\label{eq:almostfinal},
\end{align}
where  we applied the divergence theorem over $\Omega$, \eqref{eq:bc1instress} and \eqref{eq:bc2instress} in \eqref{eq:partialresult}, and the divergence theorem over $\Gamma_m$ in \eqref{eq:almostfinal} together with the fact that ${\bf v}=0$ on $\Gamma_w$, and in particular, ${\bf v}=0$ on $\overline{\Gamma_m} \cap \overline{\Gamma_w}$. The weak form of the incompressibility condition \eqref{eq:weak2} is obtained by multiplying \eqref{eq:2b} by any $q\in {\cal P}$ and integrating over $\Omega$. 

Finally, the weak compatibility between meniscus displacements and normal velocities \eqref{eq:weakcompatibility} is obtained by  computing the surface gradient $\nabla_S$ on both sides of   \eqref{eq:2c}, multiplying by a test function $\zeta\in \Xi$ and integrating over $\Gamma_m$. The compatibility condition \eqref{eq:2c} is recovered from the weak statement \eqref{eq:weakcompatibility}  by first noticing that $\lambda\hat \xi-\hat u_z\in \Xi$, and then setting $\zeta=\overline{\lambda\hat \xi-\hat u_z}$ in \eqref{eq:weakcompatibility} to conclude from \eqref{eq:prop2} that $\lambda\hat \xi=\hat u_z$ on $\Gamma_m$.

\subsection{Sign of the Real Part of the Eigenvalues of the Continuous Problem}
\label{app:eigenvaluesign}

We formally show next that for any non-trivial solution $\hat{\mathbf u},\hat{p},\hat{\xi}$ of the weak form \eqref{eq:weaks},%
\footnote{For simplicity of notation, we drop the $\hat\cdot$ notation in the rest of this section.}
the eigenvalue $\lambda$ needs to have a negative real part, i.e., $\mathfrak{Re}(\lambda)<0$. 

To see that $\mathfrak{Re}(\lambda)<0$, notice first that for a  solution $({\bf u}, p, \xi)$with ${\bf u}\not={\bf 0}$, $a( {\bf u},\overline{ {\bf u}})>0$, $c( {\bf u},\overline {\bf u})>0$ (from \eqref{eq:prop3}), and $s(\overline\xi,\xi)\ge 0$, while $0=b( {\bf u}, \overline p)=\overline{b(\overline{\bf u}, p)}$ from  \eqref{eq:weakp}. Therefore
\begin{align}
    \mathfrak{Re}\left[g \big(({\bf u}, p, \xi),(\overline {\bf u}, \overline p, -\overline \xi) \big)\right] & = c({\bf u},\overline{\bf u})+ \mathfrak{Re}\left[s(\xi, \overline{u_z})-s(u_z,\overline{\xi}) + b(\overline{\bf u},p)+ b({\bf u},\overline p)\right] \nonumber \\ 
    &= c({\bf u},\overline{\bf u}), \label{eq:reg} \\
    h \big(({\bf u}, p, \xi),(\overline {\bf u}, \overline p, -\overline \xi) \big) ~ & = -a({\bf u},\overline{\bf u})- s(\xi,\overline\xi)<0, \nonumber
\end{align}
and from \eqref{eq:symmetriceigenvaluepb}, it follows that 
\begin{equation}
    \mathfrak{Re}(\lambda) = \mathfrak{Re}\left[\frac{g(({\bf u}, p, \xi),(\overline {\bf u}, \overline p,- \overline \xi)) }{h(({\bf u}, p, \xi),(\overline {\bf u}, \overline p, -\overline \xi))}\right] = -\frac{c({\bf u},\overline{\bf u})}{a({\bf u},\overline{\bf u})+ s(\xi,\overline\xi)}<0.
\end{equation}
    
Next, if $({\bf 0}, p, \xi)$ is a solution (${\bf u}={\bf 0}$), we show that $\xi=0$ and $p=0$, so there are no non-trivial solutions of  weak form \eqref{eq:weakp}. We can then conclude that $\mathfrak{Re}(\lambda)<0$ for any eigenvalue $\lambda$.

Assuming that ${\bf u}={\bf 0}$, the fact that $\xi=0$ and $p=0$ can be obtained from \eqref{eq:symmetriceigenvaluepb} as well using the  inf-sup condition for Stokes problem (see, e.g. \cite{bramble2003proof,brezzi2012mixed}), but for the sake of simplicity we provide a formal argument here based on \eqref{eq:linearizedequationsfourier}. Since ${\bf u}={\bf 0}$, \eqref{eq:2} implies that $\nabla p=0$ in $\Omega$ and \eqref{eq:5} implies that $ p=0$ on $\Gamma_t$, from where we conclude that $ p=0$ in $\Omega$. Similarly, \eqref{eq:3} implies that $\Delta_S  \xi=0$ on $\Gamma_m$, and hence that $ \xi$ is an affine function on $\Gamma_m$ that is equal to zero on $\overline{\Gamma_m} \cap \overline{\Gamma_w}$, from where we conclude that $\xi=0$.

\subsection{Sign of the Real Part of the Eigenvalues of the Discrete Problem }
\label{app:invertibilityofG}

We prove next that if $\lambda$ is an eigenvalue of the generalized eigenvalue problem \eqref{eq:symmetriceigenvaluepb_FE}, then $\mathfrak{Re}(\lambda)<0$. In particular, this implies that the matrix $G$ is invertible. 

This result is a consequence of the fact that if $X=[U\ P\ Z]^{\sf T}$ corresponds to $({\bf u}_h,p_h,\xi_h)\in \mathcal V_h\times \mathcal P_h\times \Xi_h$ with ${\bf u}_h\neq{\bf 0}$, then as in \eqref{eq:reg}, 
\begin{align}
    \mathfrak{Re}\left[g \big(({\bf u}_h, p_h, \xi_h),(\overline {\bf u}_h, \overline p_h, -\overline \xi_h) \big)\right] & =  c({\bf u}_h,\overline{\bf u}_h)>0, \\
    h \big(({\bf u}_h, p_h, \xi_h),(\overline {\bf u}_h, \overline p_h, -\overline \xi_h) \big) ~ & = -a({\bf u}_h,\overline{\bf u}_h)- s(\xi_h,\overline\xi_h)<0.\nonumber
\end{align}
From \eqref{eq:symmetriceigenvaluepb_FE}, it follows that 
$$
\mathfrak{Re}(\lambda) = \mathfrak{Re}\left[\frac{g(({\bf u}_h, p_h, \xi_h),(\overline {\bf u}_h, \overline p_h,- \overline \xi_h)) }{h(({\bf u}_h, p_h, \xi_h),(\overline {\bf u}_h, \overline p_h, -\overline \xi_h))}\right] = -\frac{c({\bf u}_h,\overline{\bf u}_h)}{a({\bf u}_h,\overline{\bf u}_h)+ s(\xi_h,\overline\xi_h)}<0.
$$

Next, we see that if ${\bf u}_h={\bf 0}$, then $p_h=0$ and $\xi_h=0$. This result does not follow directly from that in \S\ref{app:eigenvaluesign} because an inf-sup condition for the discrete problem is needed, as we show next. 

By selecting any ${\bf v}_h\in \mathcal V_h$ such that ${\bf v}_h\cdot {\bf e}_z=0$ on $\Gamma_m$, we conclude from \eqref{eq:symmetriceigenvaluepb_FE} that 
\begin{equation}
\label{eq:binfsupfollow}
    0=g(({\bf 0},p_h,\xi_h),({\bf v}_h,0,0))= b({\bf v}_h,p_h).
\end{equation}
To conclude that $p_h=0$, we need to take advantage of inf-sup conditions \eqref{eq:infsup} in both planar and axisymmetric flows. We first show that the real and imaginary parts are constants in $\Omega$. To this end, let $p_{av}\in \mathbb C$ be the average of $p_h$, i.e. $|\Omega|p_{av}=\int_\Omega p_h\; dV$,  and let $\tilde p_h=p_h-p_{av}\in \mathcal P_{h,0}$. Notice that for ${\bf v}_h\in \mathcal V_{h,0}$,
\begin{equation}
     b({\bf v}_h,p_{av})=-p_{av}\int_{\partial \Omega}{\bf v}_h\cdot {\bf n}\; dS = 0\Longrightarrow b({\bf v}_h,p_h)=b({\bf v}_h,\tilde p_h).
\end{equation}
Next, from \eqref{eq:binfsupfollow} and \eqref{eq:infsup} we can write that for any ${\bf v}_h\in \mathfrak{Re}({\mathcal V}_{h,o})\subset\mathcal V_h$, $\mathfrak{Re}({\bf v}_h)\not={\bf 0}$,
\begin{displaymath}
    0 = \frac{\mathfrak{Re}\left[b({\bf v}_h,p_h)\right]}{\|{\bf v}_h\|_{1,2}} =\frac{\mathfrak{Re}\left[b({\bf v}_h,\tilde p_h)\right]}{\|{\bf v}_h\|_{1,2}}=\frac{b({\bf v}_h,\mathfrak{Re}(\tilde p_h))}{\|{\bf v}_h\|_{1,2}}> c \|\mathfrak{Re}(\tilde p_h)\|_{0,2},
\end{displaymath}
from where we conclude that $\mathfrak{Re}(\tilde p_h)=0$. Similarly, we conclude that $\mathfrak{Im}(\tilde p_h)=0$ and hence that $p_h=p_{av}$ from
\begin{displaymath}
    0 = \frac{\mathfrak{Re}\left[b(-i{\bf v}_h,p_h)\right]}{\|{\bf v}_h\|_{1,2}} =\frac{\mathfrak{Re}\left[b(-i{\bf v}_h,\tilde p_h)\right]}{\|{\bf v}_h\|_{1,2}}=\frac{b({\bf v}_h,\mathfrak{Im}(\tilde p_h))}{\|{\bf v}_h\|_{1,2}}> c \|\mathfrak{Im}(\tilde p_h)\|_{0,2}.
\end{displaymath}
To see that $p_{av}=0$, and conclude that $p_h=0$, it suffices to select ${\bf v}_h\in \mathcal V_h$ such that $\int_{\partial \Omega} {\bf v}_h\cdot {\bf n}\; dS\neq 0$ in \eqref{eq:binfsupfollow}; namely,
$$
0 = b({\bf v}_h,p_{av})=-p_{av}\int_{\partial \Omega}{\bf v}_h\cdot {\bf n}\; dS. 
$$
To show that $\xi_h=0$, we select  ${\bf v}\in \mathcal V_h$ such that ${\bf v}\cdot {\bf e}_z=\overline{\xi_h}$ on $\Gamma_m$, and write
\begin{equation}
\label{eq:invertforxi}
    0 = GX \Rightarrow 0=g(({\bf 0},0,\xi_h),({\bf v},0,0))= s(v_z,\xi_h)=s(\overline{\xi_h},\xi_h).
\end{equation}
Then \eqref{eq:prop2} implies that $\xi_h=0$, and hence that $X=[0\ 0\ 0]^{\sf T}$.

\subsection{Distance between Eigenvectors}
\label{sec:adef}

Next, We show that the computation of $\beta(X_1,X_2)$ in \eqref{eq:adef}. From \eqref{eq:adef} we have
\begin{displaymath}
    \begin{aligned}
        d(X_1,X_2) & = \min_{\beta\in \mathbb C,|\beta|=1} |X_1-\beta X_2|^2 \\
        & = \min_{\beta\in \mathbb C,|\beta|=1} (X_1^{\sf T}-\beta X_2^{\sf T})(\overline{X}_1-\overline{\beta}\overline{X}_2) \\
        & =\min_{\beta\in \mathbb C,|\beta|=1} 2-2 \mathfrak{Re}(\beta X_2^{\sf T}\overline{X}_1),
    \end{aligned}
\end{displaymath}
where we assumed that $X_i^{\sf T}\overline{X}_i=1$ for $i=1,2$.
The minimum is attained when $\mathfrak{Re}(\beta X_2^{\sf T}\overline{X}_1) $ is maximized, or for $\beta=\overline{X_2^{\sf T}\overline{X}_1}/|X_2^{\sf T}\overline{X}_1|$, which is \eqref{eq:adef}. Finally, by substitution we obtain
\begin{equation}
    \label{eq:distance}
    d(X_1,X_2) = 2(1 -  |X_2^{\sf T}\overline{X}_1|).
\end{equation}

%% file: main.bbl
\begin{thebibliography}{10}
\expandafter\ifx\csname url\endcsname\relax
  \def\url#1{\texttt{#1}}\fi
\expandafter\ifx\csname urlprefix\endcsname\relax\def\urlprefix{URL }\fi
\expandafter\ifx\csname href\endcsname\relax
  \def\href#1#2{#2} \def\path#1{#1}\fi

\bibitem[Amestoy {\rm et~al.}, 2011]{Amestoy2011}
Amestoy, P., Buttari, A., Duff, I., Guermouche, A., L'Excellent, J.-Y.  and
  U{\c{c}}ar, B. (2011{\rm{}}).
\newblock {\rm MUMPS. In: Encyclopedia of Parallel Computing} pp. 1232--1238.
\newblock 11 W 42nd St, New York, NY: Springer US.

\bibitem[Arafa, 2007]{Arafa2007}
Arafa, M. (2007{\rm{}}).
\newblock Finite element analysis of sloshing in rectangular liquid-filled
  tanks.
\newblock {\rm J. Vib. Control }, \emph{13}, 883--903.

\bibitem[Balay {\rm et~al.}, 2022]{Balay2022}
Balay, S. {\rm et~al.} (2022{\rm{}}).
\newblock {PETSc} ({P}ortable, {E}xtensible {T}oolkit for {S}cientific
  {C}omputation) {U}sers {M}anual, {ANL}-21/39 - {R}evision 3.17.
\newblock Technical report, Mathematics and Computer Science Division, Argonne
  National Laboratory.

\bibitem[Beale, 1984]{Beale1984}
Beale, J.~T. (1984{\rm{}}).
\newblock {Large-time regularity of viscous surface waves}.
\newblock {\rm Archive for Rational Mechanics and Analysis }, \emph{84},
  307--352.

\bibitem[Benjamin and Scott, 1970]{Benjamin1970}
Benjamin, T.~B. and Scott, J.~C. (1970{\rm{}}).
\newblock Gravity-capillary waves with edge constraints.
\newblock {\rm J. Fluid Mech.}, \emph{92}, 241--267.

\bibitem[Biswal {\rm et~al.}, 2003]{Biswal2003}
Biswal, K.~C., Bhattacharyya, S.~K.  and Sinha, P.~K. (2003{\rm{}}).
\newblock Free-vibration analysis of liquid-filled tank with baffles.
\newblock {\rm J. Sound Vib.}, \emph{259}, 177--192.

\bibitem[Boffi, 1997]{Boffi1997}
Boffi, D. (1997{\rm{}}).
\newblock Three-dimensional finite element methods for the {S}tokes problem.
\newblock {\rm SIAM J. Numer. Anal.}, \emph{34}, 664--670.

\bibitem[Bramble, 2003]{bramble2003proof}
Bramble, J.~H. (2003{\rm{}}).
\newblock A proof of the inf--sup condition for the {S}tokes equations on
  {L}ipschitz domains.
\newblock {\rm Mathematical Models and Methods in Applied Sciences },
  \emph{13}, 361--371.

\bibitem[Brezzi and Fortin, 2012]{brezzi2012mixed}
Brezzi, F. and Fortin, M. (2012{\rm{}}).
\newblock {\rm Mixed and hybrid finite element methods}, vol. 15,.
\newblock Springer Science \& Business Media.

\bibitem[Chantasiriwan, 2009]{Chantasiriwan2009}
Chantasiriwan, S. (2009{\rm{}}).
\newblock Modal analysis of free vibration of liquid in rigid container by the
  method of fundamental solutions.
\newblock {\rm Engineering Analysis with Boundary Elements } \emph{33},
  726--730.

\bibitem[Choudhary {\rm et~al.}, 2021]{Choudhary2021}
Choudhary, N., Kumar, N., Strelnikova, E., Gnitko, V., Kriutchenko, D.  and K.,
  D. (2021{\rm{}}).
\newblock Liquid vibrations in cylindrical tanks with flexible membranes.
\newblock {\rm Journal of King Saud University - Science } \emph{33}, 101589.

\bibitem[Ciarlet, 2021]{ciarlet2021mathematical}
Ciarlet, P.~G. (2021{\rm{}}).
\newblock {\rm Mathematical elasticity: Three-dimensional elasticity}.
\newblock SIAM.

\bibitem[Cocciaro {\rm et~al.}, 1993]{Cocciaro1993}
Cocciaro, B., Faetti, S.  and Festa, C. (1993{\rm{}}).
\newblock Experimental investigation of capillarity effects on surface gravity
  waves: non-wetting boundary conditions.
\newblock {\rm J. Fluid Mech.} \emph{246}, 43--66.

\bibitem[Cruchaga {\rm et~al.}, 2013]{Cruchaga2013}
Cruchaga, M., Reinoso, R., Storti, M., Celentano, D.  and Tezduyar, T.
  (2013{\rm{}}).
\newblock Finite element computation and experimental validation of sloshing in
  rectangular tanks.
\newblock {\rm Computational Mechanics } \emph{52}, 1301--1312.

\bibitem[El-Kamali {\rm et~al.}, 2011]{ElKamali2011}
El-Kamali, M., Schott{\'e}, J.-S.  and Ohayon, R. (2011{\rm{}}).
\newblock Three-dimensional modal analysis of sloshing under surface tension.
\newblock {\rm International Journal for Numerical Methods in Fluids }
  \emph{65}, 87--105.

\bibitem[Ern and Guermond, 2004]{ern2004theory}
Ern, A. and Guermond, J.-L. (2004{\rm{}}).
\newblock {\rm Theory and practice of finite elements}, vol. 159,.
\newblock Springer.

\bibitem[Farhat {\rm et~al.}, 2013]{Farhat2013}
Farhat, C., Chiu, E. K.-y., Amsallem, D., Schott{\'e}, J.-S.  and Ohayon, R.
  (2013{\rm{}}).
\newblock Modeling of fuel sloshing and its physical effects on flutter.
\newblock {\rm AIAA Journal } \emph{51}, 2252--2265.

\bibitem[Gavrilyuk {\rm et~al.}, 2006]{Gavrilyuk2006}
Gavrilyuk, I., Lukovsky, I., Trotsenko, Y.  and Timokha, A. (2006{\rm{}}).
\newblock Sloshing in a vertical circular cylindrical tank with an annular
  baffle. {P}art 1. {L}inear fundamental solutions.
\newblock {\rm J Eng Math } \emph{54}.

\bibitem[Geuzaine and Remacle, 2009]{Geuzaine2009}
Geuzaine, C. and Remacle, J.-F. (2009{\rm{}}).
\newblock Gmsh: a three-dimensional finite element mesh generator with built-in
  pre- and post-processing facilities.
\newblock {\rm Int. J. Numer. Meth. Eng.} \emph{79}, 1309--1331.

\bibitem[Graham-Eagle, 1983]{GrahamEagle1983}
Graham-Eagle, J. (1983{\rm{}}).
\newblock A new method for calculating eigenvalues with applications to
  gravity-capillary waves with edge constraints.
\newblock {\rm Math. Proc. Cambridge Philos. Soc.} \emph{94}, 553--564.

\bibitem[Hern\'{a}ndez {\rm et~al.}, 2007]{Hernandez2007}
Hern\'{a}ndez, V., Roman, J.~E., Tom\'{a}s, A.  and Vidal, V. (2007{\rm{}}).
\newblock Krylov-{S}chur methods in {SLEPc}, {STR}-7.
\newblock Technical report Universitat Polit\`{e}cnica de Val\`{e}ncia.

\bibitem[Karampelas {\rm et~al.}, 2017]{Karampelas2017}
Karampelas, I., Vader, S., Vader, Z., Sukhotskiy, V., Verma, A., Garg, G.,
  Tong, M.  and Furlani, E. (2017{\rm{}}).
\newblock Drop-on-demand 3D metal printing.
\newblock {\rm TechConnect Briefs } \emph{4}, 153--155.

\bibitem[Kashani {\rm et~al.}, 2020]{Kashani2020}
Kashani, S.~Y., Afzalian, A., Shirinichi, F.  and Moraveji, M.~K.
  (2020{\rm{}}).
\newblock Microfluidics for core-shell drug carrier particles - a review.
\newblock {\rm RSC Adv.} \emph{11}, 229--249.

\bibitem[Kidambi, 2009{\rm{a}}]{Kidambi2009a}
Kidambi, R. (2009{\rm{a}}).
\newblock Meniscus effects on the frequency and damping of capillary-gravity
  waves in a brimful circular cylinder.
\newblock {\rm Wave Motion } \emph{46}, 144--154.

\bibitem[Kidambi, 2009{\rm{b}}]{Kidambi2009b}
Kidambi, R. (2009{\rm{b}}).
\newblock Capillary damping of inviscid surface waves in a circular cylinder.
\newblock {\rm J. Fluid Mech.} \emph{627}, 323--340.

\bibitem[Kim {\rm et~al.}, 2018]{Kim2018}
Kim, S.~H., Kang, H., Kang, K., Lee, S.~H., Cho, K.~H.  and Hwang, J.~Y.
  (2018{\rm{}}).
\newblock Effect of Meniscus Damping Ratio on Drop-on-Demand
  Electrohydrodynamic Jetting.
\newblock {\rm Applied Sciences } \emph{8}.

\bibitem[Landau and Lifshitz, 1987]{LandauLifshitz1987}
Landau, L.~D. and Lifshitz, E.~M. (1987{\rm{}}).
\newblock {\rm Volume 6 of Course of Theoretical Physics: Fluid Mechanics}.
\newblock 2nd edition, Pergamon Press.

\bibitem[Langtangen and Logg, 2016]{Langtangen2016}
Langtangen, H.~P. and Logg, A. (2016{\rm{}}).
\newblock {\rm Solving PDEs in Python: The FEniCS Tutorial I}.
\newblock Springer.

\bibitem[Larson and Bengzon, 2013]{LarsonBengzon2013}
Larson, M.~G. and Bengzon, F. (2013{\rm{}}).
\newblock {\rm The Finite Element Method: Theory, Implementation, and
  Applications}.
\newblock Springer.

\bibitem[Lee and Li, 2011]{lee2011stability}
Lee, Y.-J. and Li, H. (2011{\rm{}}).
\newblock On stability, accuracy, and fast solvers for finite element
  approximations of the axisymmetric {S}tokes problem by {H}ood--{T}aylor
  elements.
\newblock {\rm SIAM journal on numerical analysis } \emph{49}, 668--691.

\bibitem[Miras {\rm et~al.}, 2012]{Miras2012}
Miras, T., Schott{\'e}, J.-S.  and Ohayon, R. (2012{\rm{}}).
\newblock Liquid sloshing damping in an elastic container.
\newblock {\rm J. Appl. Mech.} \emph{79}, 010902.

\bibitem[Nicol\'{a}s, 2002]{Nicolas2002}
Nicol\'{a}s, J.~A. (2002{\rm{}}).
\newblock The viscous damping of capillary-gravity waves in a brimful circular
  cylinder.
\newblock {\rm Phys. Fluids } \emph{14}, 1910--1919.

\bibitem[Nicol\'{a}s, 2005]{Nicolas2005}
Nicol\'{a}s, J.~A. (2005{\rm{}}).
\newblock Effects of static contact angles on inviscid gravity-capillary waves.
\newblock {\rm Phys. Fluids } \emph{17}, 022101.

\bibitem[{Norwegian University of Science and Technology (NTNU) -- Department
  of Ma-thematical Sciences}, 2013]{NTNU2013}
{Norwegian University of Science and Technology (NTNU) -- Department of
  Ma-thematical Sciences} (2013{\rm{}}).
\newblock {TMA4220} Autumn 2013 Programming project.

\bibitem[Ohayon and Schott{\'e}, 2016]{Ohayon2016}
Ohayon, R. and Schott{\'e}, J.-S. (2016{\rm{}}).
\newblock {\rm Advances in Computational Fluid-Structure Interaction and Flow
  Simulation: New Methods and Challenging Computations} chapter Modal Analysis
  of Liquid--Structure Interaction, pp. 423--438.
\newblock Cham: Springer International Publishing.

\bibitem[Ohayon and Soize, 2015]{Ohayon2015}
Ohayon, R. and Soize, C. (2015{\rm{}}).
\newblock Vibration of structures containing compressible liquids with surface
  tension and sloshing effects. {R}educed-order model.
\newblock {\rm Computational Mechanics }, \emph{55}, 1071--1078.

\bibitem[Roman {\rm et~al.}, 2022]{Roman2022}
Roman, J.~E., Campos, C., Dalcin, L., Romero, E.  and Tom\'{a}s, A.
  (2022{\rm{}}).
\newblock {SLEPc} ({S}calable {L}ibrary for {E}igenvalue {P}roblem
  {C}omputations) {U}sers {M}anual, {DSIC}-II/24/02 - {R}evision 3.17.
\newblock Technical report, Universitat Polit\`{e}cnica de Val\`{e}ncia.

\bibitem[Schweizer, 1997]{Schweizer1997}
Schweizer, B. (1997{\rm{}}).
\newblock {Free Boundary Fluid Systems in a Semigroup Approach and Oscillatory
  Behavior}.
\newblock {\rm SIAM Journal on Mathematical Analysis }, \emph{28}, 1135--1157.

\bibitem[Seemann {\rm et~al.}, 2011]{Seemann2011}
Seemann, R., Brinkmann, M., Pfohl, T.  and Herminghaus, S. (2011{\rm{}}).
\newblock Droplet based microfluidics.
\newblock {\rm Reports on progress in physics }, \emph{75}, 016601.

\bibitem[Seeto {\rm et~al.}, 2022]{Seeto2022}
Seeto, W.~J., Tian, Y., Pradhan, S., Minond, D.  and Lipke, E.~A.
  (2022{\rm{}}).
\newblock Droplet Microfluidics-Based Fabrication of Monodisperse Poly(ethylene
  glycol)-Fibrinogen Breast Cancer Microspheres for Automated Drug Screening
  Applications.
\newblock {\rm ACS Biomater. Sci. Eng.}, \emph{8}, 3831--3841.

\bibitem[Seo {\rm et~al.}, 2022]{Seo2022}
Seo, J., Somarakis, C., Korneev, S., Behandish, M.  and Lew, A.~J.
  (2022{\rm{}}).
\newblock Physics-based nozzle design rules for high-frequency liquid metal
  jetting.
\newblock {\rm Physics of Fluids }, \emph{34}, 102113.

\bibitem[Stachewicz {\rm et~al.}, 2009]{Stachewitz2009}
Stachewicz, U., Dijksman, J.~F., Burdinski, D., Yurteri, C.~U.  and
  Marijnissen, J. C.~M. (2009{\rm{}}).
\newblock Relaxation Times in Single Event Electrospraying Controlled by Nozzle
  Front Surface Modification.
\newblock {\rm Langmuir }, \emph{25}, 2540--2549.

\bibitem[Stewart, 2001]{Stewart2001}
Stewart, G.~W. (2001{\rm{}}).
\newblock A {K}rylov-{S}chur algorithm for large eigenproblems.
\newblock {\rm SIAM J. Matrix Anal. Appl.}, \emph{23}, 601--614.

\bibitem[Stewart, 2002]{Stewart2002}
Stewart, G.~W. (2002{\rm{}}).
\newblock Addendum to ``A {K}rylov-{S}chur algorithm for large eigenproblems".
\newblock {\rm SIAM J. Matrix Anal. Appl.}, \emph{24}, 599--601.

\bibitem[Taira {\rm et~al.}, 2017]{Taira2017}
Taira, K., Brunton, S.~L., Dawson, S. T.~M., Rowley, C.~W., Colonius, T.,
  McKeon, B.~J., Schmidt, O.~T., Gordeyev, S., Theofilis, V.  and Ukeiley,
  L.~S. (2017{\rm{}}).
\newblock Modal Analysis of Fluid Flows: An Overview.
\newblock {\rm AIAA Journal }, \emph{55}, 4013--4041.

\bibitem[Taylor and Hood, 1973]{Taylor1973}
Taylor, C. and Hood, P. (1973{\rm{}}).
\newblock A numerical solution of the {N}avier-{S}tokes equations using the
  finite element technique.
\newblock {\rm Comput. Fluids }, \emph{1}, 73--100.

\bibitem[Viola and Gallaire, 2018]{Viola2018}
Viola, F. and Gallaire, F. (2018{\rm{}}).
\newblock Theoretical framework to analyze the combined effect of surface
  tension and viscosity on the damping rate of sloshing waves.
\newblock {\rm Physical Review Fluids }, \emph{3}, 094801.

\bibitem[Wang {\rm et~al.}, 2006]{Wang2006}
Wang, W., Li, J.  and Wang, T. (2006{\rm{}}).
\newblock Damping computation of liquid sloshing with small amplitude in rigid
  container using {FEM}.
\newblock {\rm Acta Mech. Sinica }, \emph{22}, 93--98.

\end{thebibliography}
